\documentclass{jfm}
\usepackage{graphicx}
\usepackage{epstopdf, epsfig, bm, url}
\usepackage{color}
\usepackage{amsmath}
\graphicspath{{figures/}}

\newcommand{\Reyp}{Re_{p}}
\newcommand{\Reyn}{Re}
\newcommand{\St}{St}
\newcommand{\Kn}{Kn}
\newcommand{\Ma}{Ma}

\def\epsK{\epsilon_{\rm K}}
\def\cs{\ensuremath{c_{\text{s}}}}
\def\kf{\ensuremath{k_{\text{f}}}}
\def\urms{\ensuremath{u_{\text{rms}}}}

\renewcommand{\vec}[1]{\ensuremath{\mbox{\boldmath$#1$}}}
\newcommand{\nab}{{\bm{\nabla}}}

\newcommand{\Sec}[1]{Section~\ref{#1}}
\newcommand{\App}[1]{Appendix~\ref{#1}}

\newcommand{\Fig}[1]{figure~\ref{#1}}

\newcommand{\Figs}[2]{figures~\ref{#1} and \ref{#2}}

\newcommand{\Figp}[2]{figure~\ref{#1}({#2})}
\newcommand{\Figsp}[3]{figures~\ref{#1}({#2}) and ({#3})}

\newcommand{\Tab}[1]{table~\ref{#1}}

\newcommand{\Eq}[1]{equation~(\ref{#1})}
\newcommand{\Eqs}[2]{equations~(\ref{#1}) and~(\ref{#2})}

\newcommand{\bra}[1]{\langle #1\rangle}
\newcommand{\meanrho}{\overline{\rho}}
\newcommand{\dd}{{\rm d}}

\def\FF{\bm{F}}
\def\ff{\bm{f}}
\def\vv{\bm{v}}
\def\uu{\bm{u}}
\def\xx{\bm{x}}
\def\kk{\bm{k}}
\newcommand{\eee}{\hat{\bm{e}}}
\definecolor{awesome}{rgb}{1.0, 0.13, 0.32}

\definecolor{antiquefuchsia}{rgb}{0.57, 0.36, 0.51}

\shorttitle{Spectral characterisation of inertial particle clustering}
\shortauthor{N.E.L. Haugen, et al.}

\title{Spectral characterisation of inertial particle clustering in turbulence}

\author{
Nils E.\ L.\ Haugen\aff{1,2,3},
Axel Brandenburg\aff{1,4},
Christer Sandin\aff{1} \and
Lars Mattsson\aff{1}
 }

\affiliation{\aff{1}Nordita,
    KTH Royal Institute of Technology and Stockholm University,
    Hannes Alfv\'ens v\"ag 12, SE-10691 Stockholm, Sweden
\aff{2}SINTEF Energi A.S., Sem Saelands vei 11, 7034 Trondheim, Norway
\aff{3} Division of Energy Science, Lule\aa \, University of Technology, Lule\aa \,  971 87, Sweden
\aff{4}The Oskar Klein Centre, Department of Astronomy, Stockholm University, AlbaNova, SE-10691 Stockholm, Sweden
}

\def\pcode{\textsc{Pencil code}}

\begin{document}
\date{\today}

\maketitle

\begin{abstract}
Clustering of inertial particles is important for many types of
astrophysical and geophysical turbulence, but it has been studied
predominately for incompressible flows.
Here we study compressible flows and
compare clustering in both compressively (irrotationally) and
vortically (solenoidally) forced turbulence.
Vortically and compressively forced flows are driven stochastically either by
solenoidal waves or by circular expansion waves, respectively.
For compressively forced flows, the power spectrum of the density of inertial
particles is a useful tool for displaying particle
clustering relative to the fluid density enhancement.
Power spectra are shown to be particularly sensitive for studying
large-scale particle clustering, while conventional tools such as
radial distribution functions are more suitable for studying
small-scale clustering.
Our primary finding is that particle clustering through shock interaction is
particularly prominent in turbulence driven by spherical
expansion waves.
It manifests itself through a double-peaked distribution of spectral
power as a function of Stokes number.
The two peaks are associated with two distinct clustering mechanisms;
shock interaction for smaller Stokes numbers and the centrifugal
sling effect for larger values.
The clustering of inertial particles is associated with the formation of caustics.
Such caustics can only be captured in the Lagrangian description,
which allows us to assess the relative importance of caustics
in vortically and compressively forced turbulence.
We show that the statistical noise resulting from the limited number
of particles in the Lagrangian description
can be removed from the particle power spectra,
allowing us a more detailed comparison of the residual spectra.
We focus on the Epstein drag law relevant for rarefied gases,
but show that our findings apply also to the usual Stokes drag.
\end{abstract}


\section{Introduction}

Turbulent gas flows often harbour particles
of different sizes, for example dust grains or water droplets.
Larger particles can have significant inertia and decouple from the flow.
This can lead to local enhancements of their density, which is
generally described as clustering.
This can be important for the coalescence of particles to larger ones.
This process eventually leads to the formation of planetesimals in
planetary accretion discs \citep{1980Icar...44..172W}
or to raindrops in clouds \citep{Shaw03,GW13}. 
Another situation where particle clustering is important is when 
reactive particles are ``competing'' for the same reactant gas.
The concentration of reactant gas may then be significantly lower
within a particle cluster than outside, which yields an overall
lower reaction rate \citep{kruger17a,2018JFM...836..932H,karchniwy19}.
For this situation, which is our main interest, it is large-scale
clustering that is most important. The reason for this is that
small particle clusters, of the order of the Kolmogorov scale,
have a diffusive time scale that is shorter than the reactive time scale
by which particles are consuming the reactant. For larger clusters,
however, the particles will consume the reactant within the 
cluster faster than diffusion can transport fresh reactant to the 
cluster.

In clouds, and in most industrial applications, the compressibility of air
is rather weak, because the turbulent flow speeds are strongly subsonic.
This tends to be quite different in astrophysical flows such as those
in accretion discs around young stars and the interstellar medium,
which is continuously being fed with dust from the outflows
of stars near the end of their lives.
The driving of turbulence is fundamentally different in the
meteorological and astrophysical contexts.
Cloud turbulence is driven by convection,
which is an intrinsically vortical flow.
There is a large variety of turbulent industrial flows, but for 
the vast majority, the
turbulence is typically driven by some sort of shear, which yields
vortical flows.
The interaction between inertial particles and shocklets in such
compressible turbulent flows has been studied by \cite{Yang+14},
who found particle clustering near shocklets.
\cite{Zhang+16} also found clustering near shocklets, but also noted
clustering in regions of low vorticity for small Mach numbers due to
the centrifuge effect.
Turbulent flows with purely compressive driving are sometimes also
referred to as acoustic (irrotational) turbulence or wave turbulence.
Turbulence in the interstellar medium is driven predominately by
supernova explosions, which are intrinsically compressive flows
\citep[see, e.g.,][]{1999ApJ...514L..99K,2004RvMP...76..125M,
2005A&A...436..585D,Federrath+11,Gent13a,Gent13b,Gent20,Evirgen19}.
At larger flow speeds, however, vorticity can always
be produced by baroclinicity and shocks 
\citep{Federrath+11,DSB11,Porter+15}.

To isolate the distinctive properties of compressive and vortical
turbulence, we simulate isothermal turbulence by applying a stochastic
forcing that is either vortical or compressive.
The assumption of isothermality is often made in the context of
interstellar turbulence \citep{Stone+98, PN02, MacLow+Klessen04},
and can be motivated by short heating and cooling times.
However, this justification may well be questioned, and we therefore
regard isothermality as the simplest assumption to focus on the new
effects of compressibility.
Including physically motivated heating and cooling processes could lead
to other new effects that are not specific to compressibility.
For compressive forcing, the pressure enhancements, which are the result of
energy injection of the forcing, are completely compressive.
It is only when the resulting spherical expansion waves interact with
each other that some vorticity can be produced -- especially for
large Mach number, which is the ratio of root-mean-square (rms) velocity
to the sound speed.
Likewise, the purely vortical driving can lead to significant
compression and finite flow divergences at larger Mach numbers
\citep[see, e.g.,][]{Federrath+11,Mattsson19a}.

For purely acoustic turbulence, the energy spectra drop slightly more rapidly
with wavenumber $k$ (like $k^{-2}$) compared to the
$k^{-5/3}$ Kolmogorov spectrum for vortical turbulence \citep{KP73}.
Acoustic turbulence does not necessarily imply large Mach numbers,
because the bulk speed may well be less than the wave speed.
Owing to viscosity, purely irrotational flows cannot exist in reality
and some level of vorticity will always be generated \citep{Federrath+11}.
Therefore, we prefer to talk about compressive turbulence instead.
Our primary interest lies in the clustering of particles in these two
types of flows for small and large Mach numbers.

There are two rather different approaches to simulating
non-interacting inertial particles.
One is to model them as a pressure-less fluid, and the
other is to model them as non-interacting point particles.
In both cases, the particles interact with the gas through friction.
We refer to these two approaches as Eulerian
and Lagrangian, respectively.
Each of them has advantages and disadvantages.
A Lagrangian description is ideal for dilute systems,
but it is susceptible to statistical noise,
especially at small length scales where there are fewer particles.
This is a disadvantage of the Lagrangian approach.
An important disadvantage of the fluid description is that it cannot
describe how particles of the same type can go past each other.
This is because in the fluid description one only considers the bulk flow,
which is the average velocity of all particles of a specific type in a
small local volume.
The bulk velocity is therefore a single-valued
function of position.
In the Lagrangian description, by contrast,
one does not average, and since there are usually
several particles in every small volume,
the flow velocity of the particle phase can be multi-valued.
In particular, when particles go past each other,
we have the formation of caustics.
This implies that particles of the same size can
have opposite velocities at the same location,
creating phase-space singularities.
In the Eulerian approach, this situation would
lead to the formation of shocks -- even for dilute
particle populations.
In the Lagrangian approach, by contrast, particle
populations can go through each other without any
interaction.

Caustic formation can be an important pathway to enhanced particle
interaction \citep{WM05,2010JFM...645..497B,Gustavsson+2012,2014JFM...749..841V}.
This applies mainly to particles large enough to decouple from the carrier
fluid and this phenomenon can be the main reason why such particles interact.
At high Mach numbers, it is mainly shock interaction and compression of
the carrier fluid when shocks meet that matters \citep{Yang+14,Zhang+16}.
This is because the density increase due to compression is by far the
greatest effect.

A more commonly studied route to enhanced interaction rates is the
centrifugal effect of turbulent eddies,
which fling the particles to the edges of the eddies \citep{ML83,Maxey87,
Squires91,Eaton94,Falkovich+02,Bec03,Bec07b,Zaichik+Alipchenkov09,
Gustavsson+Mehlig11,Bragg+Collins14,Bragg+15,Bragg17,Bhatnagar18,Yavuz18}.
This is because the particles do not experience the
confining pressure that keeps the gas on closed streamlines.
The relative importance of caustics and the centrifugal
effects is not well understood \citep{2014JFM...749..841V}.
One of the goals of the present study is therefore
to assess their roles separately for vortical and
compressively forced turbulence.
This assessment will be based on the knowledge that caustics are
only captured by the Lagrangian approach while the centrifugal effect
is captured both by the Lagrangian and the Eulerian approaches.

The Lagrangian and Eulerian descriptions are complementary
and can also be used to gauge their respective regimes of validity.
This allows us to study, for example, when and at what
length scales statistical noise becomes important,
and when caustics formation becomes important.
We focus on three-dimensional simulations, but we
also use one-dimensional simulations, where caustic formation
can be studied in isolation.

In both types of approaches, we ignore the back-reaction
of particles on the flow.
This can become important at large mass loading parameters
and can lead to other interesting effects such as the
streaming instability \citep{JY07J} and the resonant drag instability 
\citep{Squire17}, which will not be addressed here.
We also neglect gravity and tidal forces.

To analyse particle clustering in incompressible turbulence, radial
distribution functions (RDF) have commonly been used
\citep{Sundaram+Collins97,Reade+Collins00,Wang+00,Salazar2008}.
They have also been used in the context of compressible transonic
turbulence \citep{Pan11}.
Alternatively, one can use a spectral approach by calculating
power spectra of particle densities.
In the context of particle clustering, we only know of the work
of \cite{2018JFM...836..932H}, who have used power spectra of
particle densities.
This approach may be more suitable for characterising particle
clustering at different length scales, including, in particular,
scales larger than the Kolmogorov scale.
Similar spectral quantities are known as structure factors in the
context of crystallography \citep{1968JChPh..48.5048J}, liquid metals
\citep{Ashcroft+Lekner66}, and biomolecular systems \citep{Essmann+95}.
The RDFs may be regarded as the real-space equivalents of these various
spectral techniques; see the work of \cite{Shaw+02} which showed
that these different measures can be related to each other;
see also the textbook of \cite{McQuarrie}.
However, they can also be complementary to each other, as we shall show
in this paper.

Contrary to earlier work on particle clustering, we are here interested
in clustering at all scales, and not just the Kolmogorov scale.
This seems particularly clear for particle clustering near shocklets,
but may in fact also be true for inertial range clustering
\citep{2018JFM...836..932H}, which is due to classical vortex
clustering at larger scales.

\section{The model}
\label{sec:model}

We consider an isothermal gas where the pressure is proportional
to the density $\rho$ and is given by $\rho\cs^2$, with $\cs$ being
the isothermal sound speed.
The velocity of the gas $\bm{u}$ is governed by the Navier-Stokes
and continuity equations,
\begin{equation}
\label{momentum}
\frac{\partial\uu}{\partial t}+\uu\cdot\bm{\nabla}\uu=
-\cs^2\bm{\nabla}\ln\rho + \bm{f}
+\rho^{-1}\bm{\nabla}\cdot(2\rho\nu\bm{\mathsf{S}}+\rho\zeta_{\rm shock}\bm{\mathsf{I}}\nab{\bm\cdot}\uu),
\end{equation}
\begin{equation}
\frac{\partial\ln\rho}{\partial t}+\uu\cdot\bm{\nabla}\ln\rho
=-\bm{\nabla}\cdot\uu,
\end{equation}
where $\bm{f}$ is a stochastic forcing term,
$\nu$ is the kinematic viscosity,
$\bm{\mathsf{I}}$ is the unit matrix with indices ${\sf I}_{ij}=\delta_{ij}$, and
$\bm{\mathsf{S}}$ is the trace-less rate of strain tensor with the components
\begin{equation}
\mathsf{S}_{ij}={\textstyle\frac{1}{2}}(\partial u_i/\partial x_j+\partial u_j/\partial x_i)
-{\textstyle\frac{1}{3}}\delta_{ij}\bm{\nabla}\cdot\bm{u}.
\end{equation}
The forcing term consists either of random plane waves (vortical forcing)
that are $\delta$-correlated in time
\citep{2012PhFl...24g5106H},
or of localised pressure enhancements (compressive forcing)
with $\bm{f}=-\bm{\nabla}\phi$,
where $\phi$ is a Gaussian in space at new locations
in regular time intervals $\delta t_{\rm f}$ \citep{2006MNRAS.370..415M}.
The amplitude of the forcing is denoted by $f_0$.
For further details, we refer the reader to \App{ForcingAlgorithms}.

In most of the simulations presented here, we perform direct numerical
simulations (DNS) in the sense that we solve the equations as stated.
In those cases, $\zeta_{\rm shock}=0$.
However, to save resources, especially in astrophysics, one often uses
a shock-capturing viscosity \citep{vNR50}.
This broadens the shocks and allows one to resolve them on a coarser mesh.
To assess the effect of such an artificial treatment on the particle
clustering, in some cases we compare the results from the DNS
with runs where a coarser mesh is used together with a shock-capturing
viscosity.
We adopt the shock-capturing viscosity of \cite{vNR50}, which corresponds to
a bulk viscosity with
\begin{equation}
\zeta_{\rm shock}=C_{\rm shock} \delta x^2 \langle-\nab{\bm \cdot}\uu\rangle_+.
\label{zeta_shock}
\end{equation}
Here, $\langle...\rangle_+$ denotes a running five point average
over all positive arguments, corresponding to a compression; see
\cite{2001A&A...369..706C} and \cite{10.1111/j.1365-2966.2004.08127.x}
for a detailed description.
In contrast to the DNS, we refer to those simulations as large eddy
simulations (LES).

It is important to realise that our Reynolds numbers are small
compared with the many types of compressible flows occurring in nature.
Therefore, our simulations are not DNS in a strict sense.
Based on numerical considerations, the kinematic viscosity cannot be
chosen too small.
Therefore, we keep its value constant, which implies that the dynamic
viscosity is enhanced in high density regions.
On physical grounds, the dynamic viscosity tends to be more nearly
constant, which would imply an enhanced kinematic viscosity in the
regions of lower density outside shocks.
This would have reduced the maximum permissible Reynolds number even
further, and might have deprived us from finding effects related to
higher Reynolds numbers.

In both the Lagrangian and Eulerian descriptions,
the velocity $\vv_p$ of the particle with
index $p$ couples to the gas through the friction force,
\begin{equation}
\label{drag}
\FF_p=-\frac{1}{\tau_p}\left(\vv_p-\uu\right).
\end{equation}
It is assumed that the particles are smaller than the 
smallest turbulent eddies, 
which have sizes that are comparable to the Kolmogorov scale.
For the flows considered here,
the particle response time is given by a term
that is slightly different for dense and dilute gases.
When the mean-free path of the gas molecules
is short, the response time is given by the
Stokes time, modified by a Reynolds number-dependent factor of the form
\begin{equation}
\label{tau_St}
\tau_p^{\text{St}}=\frac{2}{9}\, \frac{\rho_p}{\rho} \, \frac{a_p^2}{\nu}
\left(1+0.15\,\Reyp^{0.687}\right)^{-1},
\end{equation}
where $a_p$ and $\rho_p$ are the radius and material density, respectively,
and Re$_{p} = a_{p}u_{\rm rms}/\nu$ is the particle Reynolds number.
\citet{Salazar2008} used a similar approach to show that the
results of their DNS were comparable to their experimental results of
particle clustering in isotropic turbulence to within the limits of
experimental uncertainty.
For rarefied gases, the response time is based on the Epstein drag and is given by
\citep{Schaaf63,Kwok75,Draine79b,2019MNRAS.490.5788M}
\begin{equation}
\label{tau_Ep}
\tau_p^{\text{Ep}}=\sqrt{\frac{\pi}{8}}\frac{\rho_p}
{\rho}\,\frac{a_p}{\cs}\left(1+\frac{9\pi}{128}
\frac{\left|\vec{u}-\vec{v}_p\right|^2}{c_{\text{s}}^2}\right)^{-1/2}.
\end{equation}
The second term inside of the parenthesis becomes important
at large Mach numbers.

In the Lagrangian description, the evolution of a particle $p$
with velocity $\vv_p$ at position $\xx_p$ is given by
\begin{equation}
\label{dvdt_dxdt}
\frac{d\vv_p}{dt}=\FF_p,\quad
\frac{d\xx_p}{dt}=\vv_p,
\end{equation}
where we have ignored the effects of Brownian motion,
which leads to the diffusion of particles.
In the associated Eulerian description, diffusion is included and,
instead of \Eq{dvdt_dxdt}, we solve instead
\begin{equation}
\label{momentump}
\frac{\partial\vv_p}{\partial t}+\vv_p\cdot\bm{\nabla}\vv_p=\FF_p
+\frac{2}{\rho_p}\bm{\nabla}\cdot(\rho_p\nu_p\bm{\mathsf{S}}_p),
\end{equation}
\begin{equation}
\label{densityp}
\frac{\partial n_p}{\partial t}+\vv_p\cdot\bm{\nabla} n_p
=-n_p\bm{\nabla}\cdot\vv_p+\kappa_p\nabla^2 n_p,
\end{equation}
where $n_p$ is the particle number density and
$\nu_p$ and $\kappa_p$ are artificial viscosity and diffusivity
for the particle fluid (denoted now by $p$ collectively for all particles).
Those terms are needed for reasons of numerical stability.

We use medium-resolution ($N_{\rm mesh}^3=256^3$ and $512^3$) in
three-dimensional (3D) triply periodic cubic domains with side lengths $L$
and the number of mesh points in each direction being $N_{\rm mesh}$.
The smallest wavenumber in the domain is then $k_1=2\pi/L$.
Unless otherwise specified, dust particles are
included as inertial particles in 5 size bins with $4\times10^6$ particles
in each (for the Lagrangian simulations).
We use the \pcode\ \citep{2020_JOSS}, which is a
high-order, finite-difference code (sixth order in space and third
order in time); see also \citet{2002CoPhC.147..471B} for details. 

We sometimes also give the dimensional values
of $f_0$, $\delta t_{\rm f}$, $\nu$, etc.
Those are based on our choice $\cs=2k_1=\bra{\rho}=1$ in the
numerical calculations.
In all cases, we use $\rho_{p}=10^3$.

\section{Diagnostic tools}
\subsection{Non-dimensional numbers}
\label{NonDimensionalNumbers}

The flow is characterised by the Reynolds and Mach numbers,
\begin{equation}
\label{eq:Reynolds}
\Reyn=\frac{\urms}{\nu\kf}
\quad\mbox{and}\quad
\Ma=\frac{u_{\rm rms}}{c_{\rm s}},
\end{equation}
respectively, where $\kf$ is the forcing wavenumber.
We also give the value of the Taylor microscale Reynolds number,
$\Rey_\lambda=\lambda u_{1D}/\nu$, which is based on the
Taylor microscale $\lambda=\sqrt{15\nu/\epsK}$ and the one-dimensional
rms velocity, $u_{1D}=\urms/\sqrt{3}$.
The behaviour of the particles is characterised by the Stokes numbers
\begin{equation}
\label{eq:stokes}
\St_{\rm int}=\tau_{p}/\tau_{\rm f}
\quad\mbox{and}\quad
\St_{\rm Kol}=\tau_{p}/\tau_{\rm Kol},
\end{equation}
where $\tau_p$ is the particle response time, 
$\tau_{\rm f}=(u_{\rm rms}\kf)^{-1}$ is the time scale related to
the size of large-scale fluid structures, e.g., the forcing scale,
and $\tau_{\rm Kol}=\sqrt{\nu/\epsK}$ is the Kolmogorov time,
where $\epsK=\bra{2\rho\nu\bm{\mathsf{S}}^2}$ is the energy dissipation rate.
Both variants of the Stokes number are related to particle clustering
due to particle inertia.
For small Mach numbers, $\rho$ is close to the mean density $\meanrho$,
allowing us to express $\epsK$ also in terms of the energy spectrum $E(k)$.
They are normalised such that $\int E(k)\,\dd k=\meanrho\bra{\uu^2}/2$.
We then have $\epsK=2\meanrho\nu\int k^2 E(k)\,\dd k$.
Also of interest is the associated Kolmogorov scale
$\ell_{\rm Kol}=(\nu^3/\epsK)^{1/4}$.

In the present work, we have defined $\tau_{\rm f}$ in terms of the
wavenumber as $(u_{\rm rms}\kf)^{-1}$.
An alternative definition would be in terms of the length scale
$2\pi/\kf$, which would make $\tau_{\rm f}$ larger by a factor of
$2\pi$, and $\St_{\rm int}$ smaller by the same factor.
This might be more meaningful, because it would 
result in a better representation of the actual separation between the 
Kolmogorov and integral scales, and hence a more correct ratio 
between $\St_{\rm int}$ and $\St_{\rm Kol}$.
We should keep this in mind when comparing these numbers in the rest of
the paper.

The Knudsen number of particles of a certain size is defined as the ratio
of the mean free path $\lambda$
of the gas molecules to the size of the particle, i.e.,
$\Kn=\lambda/d_{\text{p}}$.
The drag force on the particles is inversely proportional to the
particle response time; see Eq.~(\ref{drag}).
For a continuous fluid, where the Knudsen number is much smaller than unity,
the particle response
time is given by the Stokesian time; see Eq.~(\ref{tau_St}).
For rarefied gases, however, the mean free path is large compared to the
particle size and the response time is then given by the
Epstein time, as given in Eq.~(\ref{tau_Ep}).

The dimensionless particle size parameter
\citep[see][]{2016MNRAS.456.4174H} is defined as
\begin{equation}
    \alpha = {\rho_{p}\over \langle\rho\rangle} {a_{p}\over L_{\rm f}},
\end{equation}
where $L_{\rm f}$ is taken to be the physical forcing scale of the
turbulent flow.
For small values of $\Kn$, we find that the mean Stokes number is
\begin{equation}
    \langle\St_{\rm int}\rangle = {\langle \tau_p^{\rm St}\rangle\over \tau_{\rm f}}
    \approx {(2/9)\,\alpha\,Re_p\over 1+0.15\,Re_p^{0.687}} \sim \alpha\,Re_p,
\end{equation}
while in the Epstein limit we find
\begin{equation}
    \langle\St_{\rm int}\rangle = {\langle\tau_p^{\rm Ep}\rangle\over \tau_{\rm f}} \approx \sqrt{\pi\over 8}\,\alpha\,\mathcal{M}_{\rm rms}\, \left(1+\frac{9\pi}{128}
\frac{\left|\vec{u}-\vec{v}_p\right|^2}{c_{\text{s}}^2}\right)^{-1/2}\sim \alpha\,\mathcal{M}_{\rm rms}.
\end{equation}
From these two relations one can see that, while $\St_{\rm int}$
for a given particle size in the Epstein limit is mainly affected by
compression and essentially unaffected by viscosity of the carrier fluid,
it is inversely proportional to the viscosity in the Stokes limit.

\subsection{Power spectra of particle density}
\label{PowerSpectra}

To measure preferential clustering at all scales, from the smallest
scale resolved in the simulation to the size of the simulation box,
we compute power spectra of $n_p$ as \citep{2018JFM...836..932H}
\begin{equation}
   P_{n}(k)=\frac{1}{2}\!\!\!\!\!\!\!\!\!\!\!\!
\sum_{k-\delta k/2 \leq |\vec{k}|<k+\delta k/2} \!\!\!\!\!\!\!\!\!\!\!\!
\left|\hat{n}_p(\vec{k})\right|^2 \text{d}^3\vec{k},
\end{equation}
where $\hat{n}_p(\vec{k})={\cal{F}} \left( n_p(\vec{x}) \right)$ is the
Fourier transform of $n_p$, $\vec{k}=(k_x,k_y,k_z)$ is the wavevector,
and the integration is over concentric shells in wave number space.
From the above, we see that
\begin{equation}
\label{eq:int_P}
\int_{k_1}^{k_{\rm max}} P_{n}(k)\,\dd k=\frac{1}{2}\langle n_p^2\rangle,
\end{equation}
where $k_{\max}=k_1 N_{\rm mesh}/2$ is the Nyquist wavenumber, and angle
brackets, $\langle...\rangle$, represent spatial averaging, which means
that $\langle n_p^2\rangle^{1/2}$ is the root-mean-squared particle
number density.

If the particles were randomly distributed, $|\hat{n}_p(\vec{k})|$
would be on average independent of $k=|\vec{k}|$.
In three dimensions, however, the shell integration introduces an
additional $k^2$ factor, so we expect
\begin{equation}
\label{eq:P_n}
    P_{n}(k)=A k^2,
\end{equation}
which can be combined with \Eq{eq:int_P} to yield
\begin{equation}
\label{eq:np}
  \frac{1}{2} \langle n_p^2\rangle
  = \int_{k_1}^{k_{\rm max}} A k^2\, \dd k
  = \frac{A}{3}(k_{\rm max}^3-k_1^3)
  \approx \frac{A}{3} k_{\rm max}^3,
\end{equation}
where we have assumed $k_1\ll k_{\max}$, and $A$ is a constant
that we shall be concerned with later in \Sec{SpectralAnalysis}.
We improve on this description further below, when we analyse
concrete examples.

\subsection{Definition of the RDFs}

To put our results into the context of other commonly used tools of
characterising particle clustering, we compare the results from our
spectral analysis with the corresponding RDFs.
They are defined as \citep{Sundaram+Collins97, Reade+Collins00,
Wang+00,Salazar2008}
\begin{equation}
g(r_i)=\frac{N_i}{N}\left/\frac{4\pi r_i^2\, \delta r}{4\pi R^3/3},\right.
\end{equation}
where $N_i$ is the number of particle pairs separated by a distance
$r_i\pm\delta r/2$, 
$N=N_p(N_p-1)/2$ is the total number of particle pairs,
$N_p$ is the number of particles
and $R$ is the largest radius that fits into the domain.

\section{Results}

\begin{figure}
\centering\includegraphics[width=0.49\textwidth]{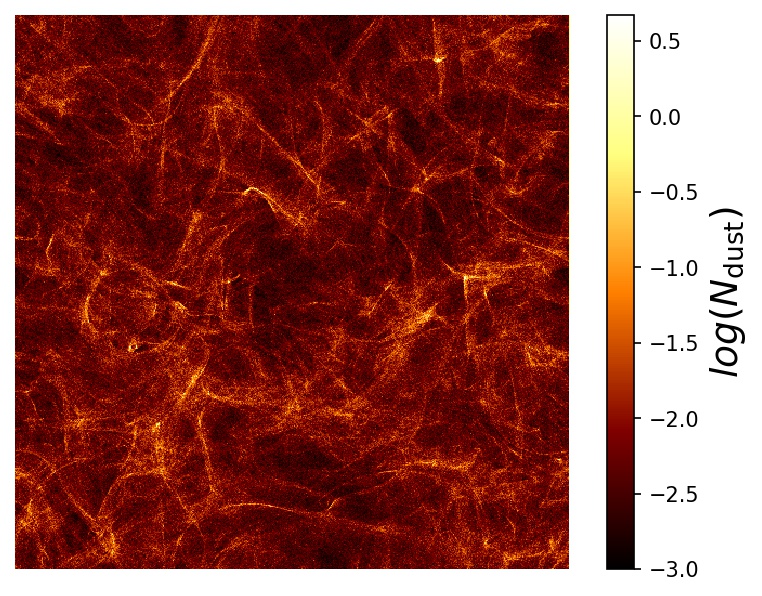}
\includegraphics[width=0.49\textwidth]{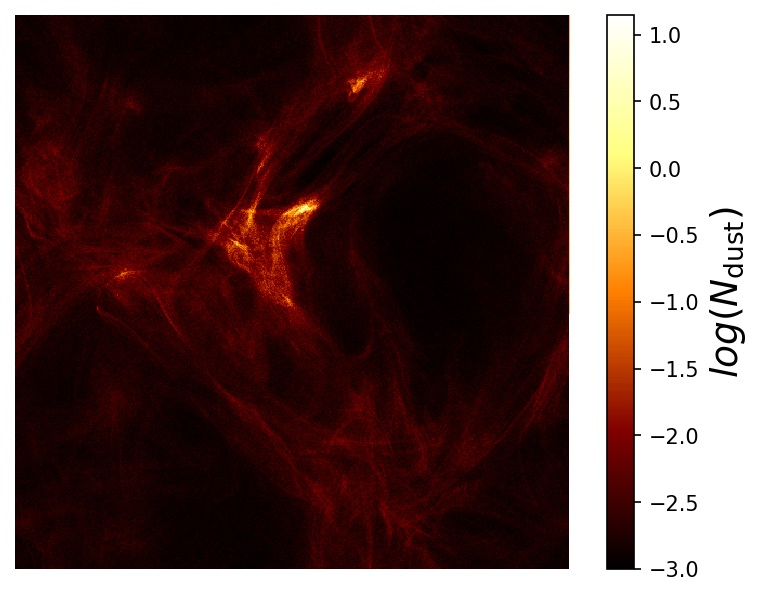}
\caption{Left: Projected particle number density in a snapshot from a DNS
with vortical forcing (later referred to as Run~V2).
Right: the same from a DNS with compressive forcing (later referred to as Run~C2).
Both cases correspond the particle size showing the most
clustering ($\St_{\rm int}=0.31$ and $0.36$, respectively).
\label{fig:projections}}
\end{figure}

In this paper, we are concerned with the differences in
particle clustering between vortical and compressive forcings.
To convey an impression of this phenomenon, we begin by showing in
\Fig{fig:projections} the projected particle number densities of snapshots
from DNSs with vortical and compressive forcings for Stokes numbers
around unity.
Evidently, the visual impressions for the two types of flow are rather
different.
Even though the typical length scales of the forcings are similar
in the two cases, the clustering phenomenon is markedly different.
For vortical forcing, the overall contrast between minimum and maximum
particle concentrations is much smaller than for compressive forcing.
In the vortical case, the particle concentrations take a more filamentary
and perhaps sheet-like structure, while in the compressive case, the
particle concentrations are more spherical in shape.

\subsection{One-dimensional simulations}

To illustrate the effects specific to compressive clustering, let us
consider first a one-dimensional shock model as an illustrative example.
Here, we also compare the Lagrangian particle simulations with the ones
in the Eulerian description.

\subsubsection{Applicability of the Eulerian approach}
\label{Applicability}

Caustics formation in the particle distribution, 
which is evident from the presence of multi-valued particle velocities, 
is a phenomenon that cannot be described with the Eulerian approach;
see \cite{Boffetta+07} and \cite{Shotorban+09} for detailed comparisons.
For small enough particle inertia, i.e., for small Stokes
numbers, the Eulerian and Lagrangian approaches should agree with
each other.
However, there is also another source of discrepancy.
The Lagrangian approach is suitable for modelling dilute systems, but,
due to the finite number of particles,
it also has the disadvantage of suffering from statistical fluctuations when
the intention is to model non-dilute systems, where fluctuations should
be small.
Statistical noise does not occur in the Eulerian approach.
Thus, for dense systems and small Stokes numbers, the Eulerian approach
can be beneficial.
To determine the limits of applicability of the Eulerian approach
quantitatively, we consider a simple one-dimensional model.

\begin{figure}
\centering
\includegraphics[width=\textwidth]{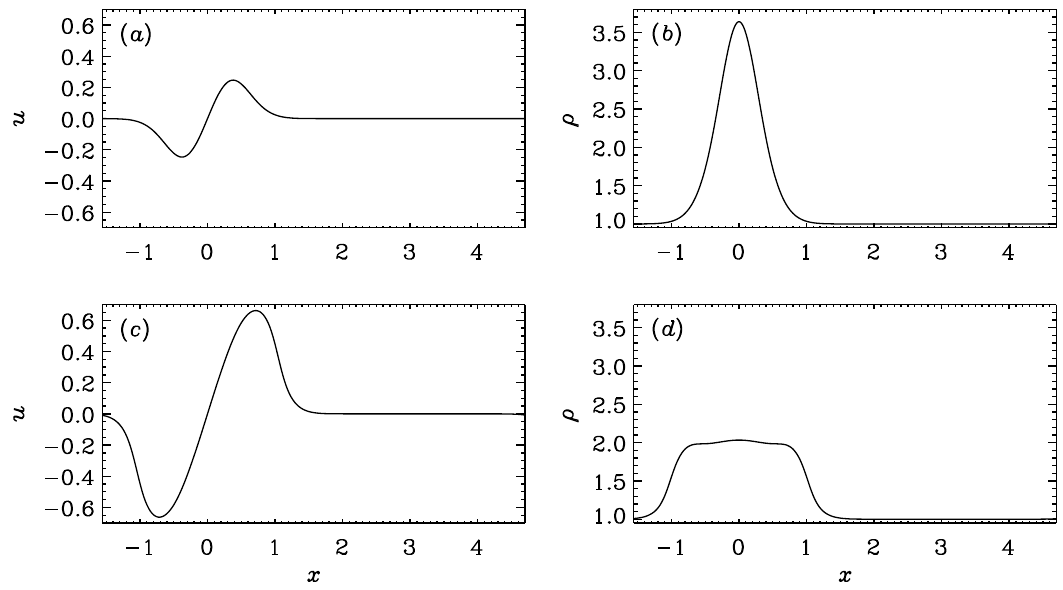}
\caption{
({\em a}) Gas velocity and ({\em b}) density at $t=0.1$ (upper row)
and $t=0.5$ (lower row) in panels ({\em c}) and ({\em d}).
The data for $t=0.5$ serve as initial condition for the particles.
\label{fig:pcomp_1d_initial}}
\end{figure}

We adopt a localised hump in the fluid density, which we model
by a Gaussian in $\ln\rho(x)$ at the position $x=0$ in a domain
of size $-\pi/2<x<3\pi/2$.
Thus, we take
\begin{equation}
    \ln\rho[(x)/\rho_0]={\cal A} \exp \left(-x^2/2\sigma_s^2 \right),
\end{equation}
where ${\cal A}$ is an amplitude factor,
$\rho_0$ is an overall normalisation coefficient,
the width is given by $\sigma=0.35$, and the ratio of 
the peak value over the background is 3.1.
For these simulations we use periodic boundary conditions.
The initial density profile launches an acoustic wave; see
\Fig{fig:pcomp_1d_initial} for plots of $u_x$ and $\rho$ at $t=0.1$
and $t=0.5$.
The front speed exceeds the sound speed when the gas speed approaches
a certain fraction of the sound speed; see Appendix~\label{pposition}
for an illustration.
We then use the gas velocity and gas density at $t=0.5$ as initial
condition for the particles by setting the particle velocity for all
particle sizes equal to the fluid velocity.
The particle number density of all particle sizes is set proportional
to the gas density.

\begin{figure}
\centering
\includegraphics[width=\textwidth]{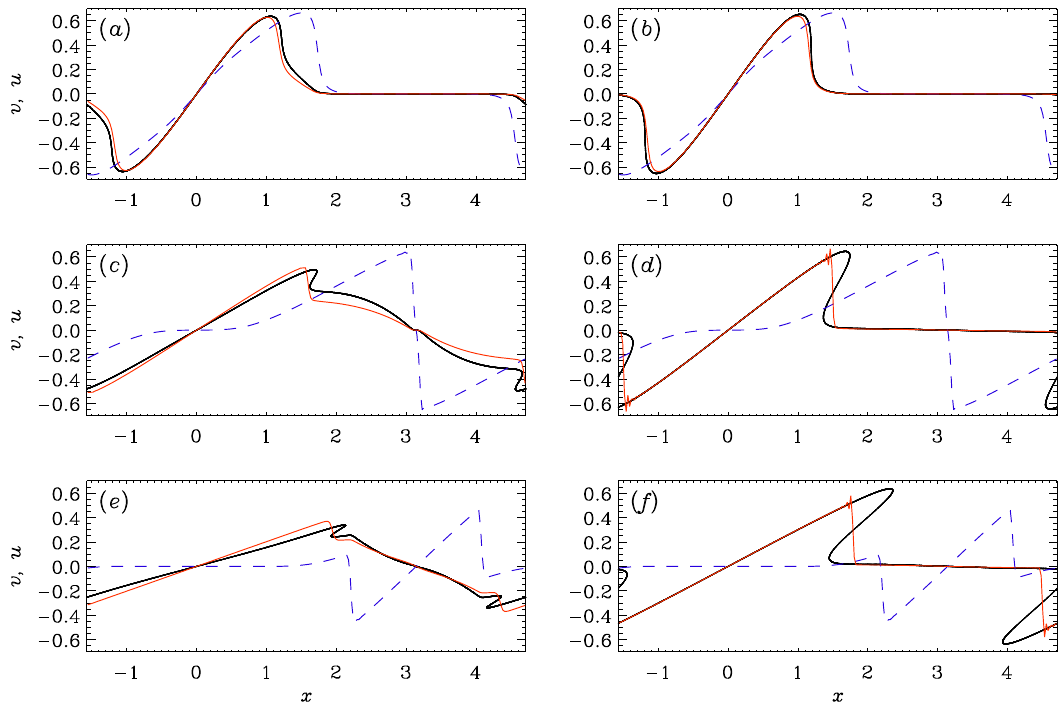}
\caption{
Particle velocity in the Lagrangian simulation (solid black) and the Eulerian
one (solid red), together with the gas velocity (dashed blue) at times $t=1$, 2,
and 3 for $a_p=3\times10^{-3}$ (corresponding to $\St_{\rm int}=1$, panels a, c, and e)
and $10^{-1}$ (corresponding to $\St_{\rm int}=30$, panels b, d, and f).
\label{fig:pcomp_1d_time}}
\end{figure}

In \Fig{fig:pcomp_1d_time} we show the particle velocities as a function
of position for Lagrangian and Eulerian models for particle radii
$a_p=3\times10^{-3}$ and $10^{-1}$ at different times.
We also show the gas velocity, which propagates at a speed slightly faster
than the sound speed, $\cs=1$; see \App{pposition} for a plot showing
the numerically obtained dependence of the front speed on the gas speed.
The lighter particles follow the fluid and are not shown, but the heavier
ones lag behind because they only inherit the speed of the gas at $t=0.5$,
and they are too heavy to get accelerated by the passing acoustic wave.
At early times ($t=1$), the velocities in the Lagrangian and Eulerian
models are close to each other, but at later times the particle velocities
in the Lagrangian description become multi-valued, which corresponds to
caustics formation.
This phenomenon becomes more prominent for the heavier particles
since they are not decelerated by the drag from the gas.
In the Eulerian description, we instead see the formation of a shock.
Away from the shock, the Lagrangian and Eulerian descriptions agree with
each other rather well, especially for the heavier particles.
To resolve the shock in the Eulerian simulations,
we must apply a certain amount
of artificial viscosity and diffusivity for the particle fluid.
If this artificial viscosity is too small, wiggles occur in the
downstream part of the shock, as can already be seen from the
profile of $v_x(x)$ in \Fig{fig:pcomp_1d_time}.
Including such artificial viscosity and diffusivity is a purely numerical
device to stabilise the solution, but it is likely to introduce errors
in the results.
By comparing with the Lagrangian approach, we will try to assess
the extent of such artifacts.

\begin{figure}
\centering
\includegraphics[width=\textwidth]{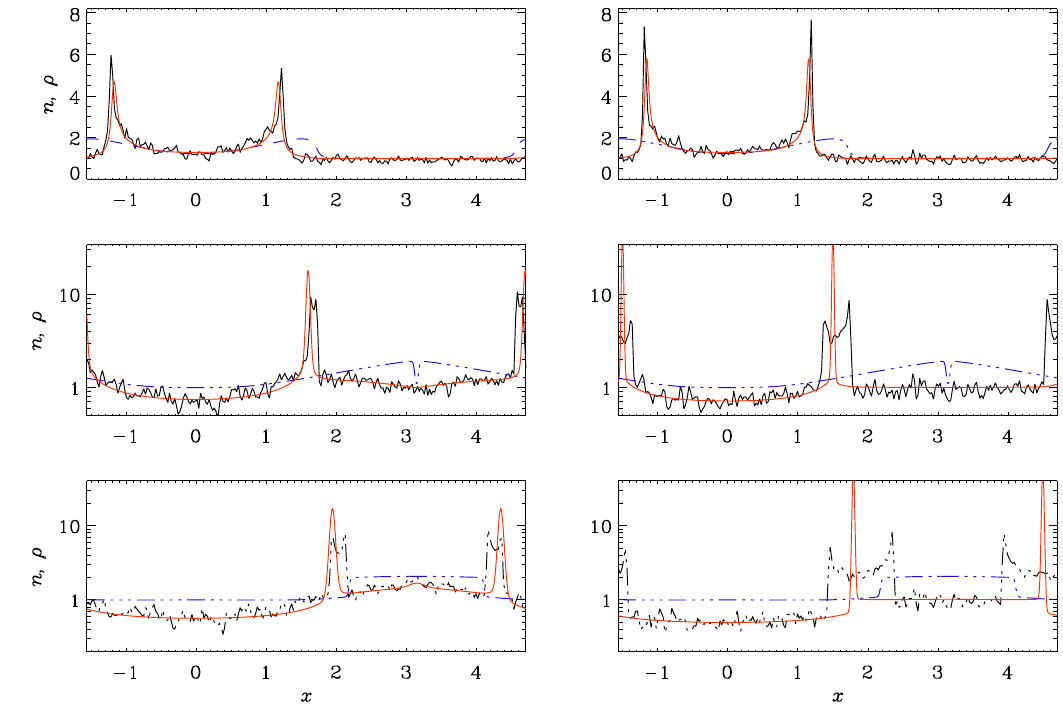}
\caption{
Particle density in the Lagrangian simulation
(solid black) and the Eulerian one (solid red), together with
the gas density (dashed blue) at times $t=1$, 2, and 3.
Two particle sizes are shown: $a_p=3\times10^{-3}$
(left column) and $10^{-1}$ (right column).
\label{fig:pcomp_1d_dens}}
\end{figure}

Snapshots of the particle number densities are shown in
\Fig{fig:pcomp_1d_dens} for the same times and the same two particle
sizes as in \Fig{fig:pcomp_1d_time}.
We see the development of extended structures
with two enhancements on their flanks, characteristic of caustics,
as is correctly reproduced with the Lagrangian approach.
The Eulerian approach, on the other hand, yields just a single albeit
very strong spike, 
which may cause an increase of the particle-interaction rate even without any caustics forming.

It is of interest to determine the Stokes number relevant for
caustics formation.
This is important for knowing the maximum Stokes number for which the
Eulerian approximation can still be used.
For smaller Stokes numbers,
no artificial particle viscosity and diffusivity are needed.
For larger Stokes numbers, however, the Eulerian approach can represent
the caustics only as shocks, which requires an increasing amount of
artificial viscosity and diffusivity to keep them numerically resolved.
The Stokes number is defined through \Eq{eq:stokes}.
For the solution shown in \Fig{fig:pcomp_1d_time}, we find that
the fluid travel time across the width of the front $\Delta x$ is
\begin{equation}
\tau_{\rm f}=\Delta x/\Delta u
\end{equation}
where $\Delta u$ is the fluid velocity at the front.
For the current experiments, $\Delta x$ is taken to be
the thickness of the front at the time 
when the particles were introduced in the simulation;
see Run~A in \Tab{tab:Stokes} for more details about the particles.

In the example discussed above, we have $\tau_{\rm f}=0.72/0.66\approx1.1$,
and we find caustics for $a_{p}\ge10^{-3}$,
which corresponds to $p=3$.
To compute the critical Stokes number $\St_{\rm crit}$
above which caustics formation occurs,
we need to know the $\rho_{p}/\rho$ ratio, where we take the fluid density
on the upstream side of the front, which is here $\rho\approx1.8$.
We used $\rho_{p}=10^3$, so, altogether, we have $\St_{\rm crit}=0.30$.
This means that, for simulations where the Stokes number is
larger than $\sim 0.3$, the Eulerian particle approach cannot be used.
It is important to realise that we are here talking about the Stokes
number based on the largest fluid scale, $\St_{\rm int}$, and not
the Kolmogorov scale.
As we shall see below, at the numerical resolutions accessible in our
three-dimensional simulations, the difference between $\St_{\rm int}$ and
$\St_{\rm Kol}$ is not very large.

\begin{table}
    \centering
    \begin{tabular}{lcccccccc}
    & $p$       & 1        &       2          &      3    &      4          &     5      &       6          & 7 \\
Run & $k_1 a_p$ & $10^{-4}$ & $3\times10^{-4}$  &  $10^{-3}$& $3\times10^{-3}$ &  $10^{-2}$ &  $3\times10^{-2}$ &  $10^{-1}$ \\
    \hline            
A & $\St_p$     & $0.03 $   & $0.1$            &  $0.3$   & $1.0$             &  $3$       &  $10$          &  $30$ \\
B & $\St_p$     & $0.007$   & $0.02$           &  $0.07$  & $0.2$             &  $0.7$     &  $2.0$         &  $6.8$ \\
    \end{tabular}
    \caption{
    Stokes numbers for Runs~A (\Sec{Applicability}) and B (\Sec{sec:1D_B}).
}
    \label{tab:Stokes}
\end{table}

\subsubsection{A mechanism for compressive supersonic clustering}
\label{sec:1D_B}

The preferential particle concentrations near shocklets in compressible
turbulent flows found and discussed by \cite{Yang+14} and \cite{Zhang+16}
suggests that irrotational supersonic flows can yield new ways
of clustering that would not occur in vortical subsonic flows.
One idea for such a mechanism is that particles of a suitable mass that
move toward each other on two colliding shocks, will be decelerated as the
shocks collide, but the particles are too heavy to become re-accelerated
as the shocks depart again immediately after the collision.
The particles will then be left behind after the shocks move away,
forming a cluster.
To test this idea, we use an experiment similar to that described in
\Sec{Applicability}, but with a stronger density enhancement.

We emphasise that the particle acceleration is always due to
the drag on the particles because of the relative velocity
difference between the particles and the fluid.
However, the drag becomes stronger when the density is high;
see \Eqs{tau_St}{tau_Ep}, so the density also plays a role.

In \Sec{Applicability}, we used the gas velocity and density at a certain
time to reinitialise the particles by setting the particle velocity for
all particle sizes to a value equal to the fluid velocity.
The initial width of the density distribution is again $0.35$.
However, the density enhancement of the gas is now so strong that its
distribution is so different from a Gaussian that it can no longer
be used for reinitialising the particles in a simple way.
We therefore reinitialise the fluid density equal to the density
of the lightest particles and then set the density and velocity of the
heavier particles to the same as the lightest particles.
The ratio of the peak value of the density over its background then
turns out to be 22.
In this experiment, we only use the Lagrangian approach.
See Run~B in \Tab{tab:Stokes} for more details about the particles.

\begin{figure}
\centering
\includegraphics[width=\columnwidth]{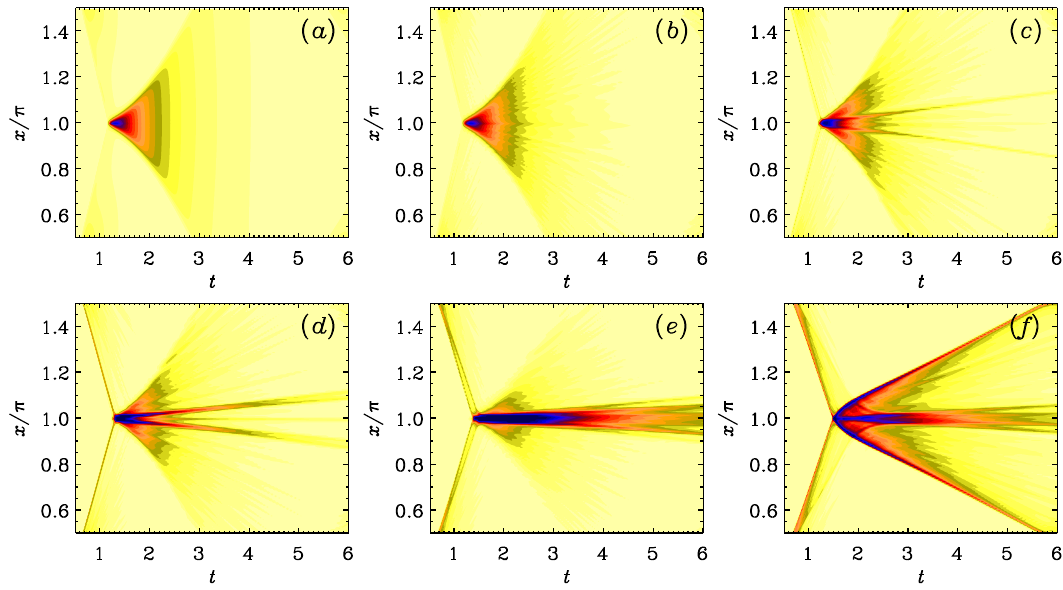}
\caption{$xt$ diagrams of ({\em a}) $\rho$, ({\em b}) $n_1$, ({\em c}) $n_4$,
({\em d}) $n_5$, ({\em e}) $n_6$, and ({\em f}) $n_7$.
Dark shades indicate high densities.
Note that shock clustering is most evident in panel~(e).
}\label{fig:pbut}
\end{figure}

\begin{figure}
\centering
\includegraphics[width=\columnwidth]{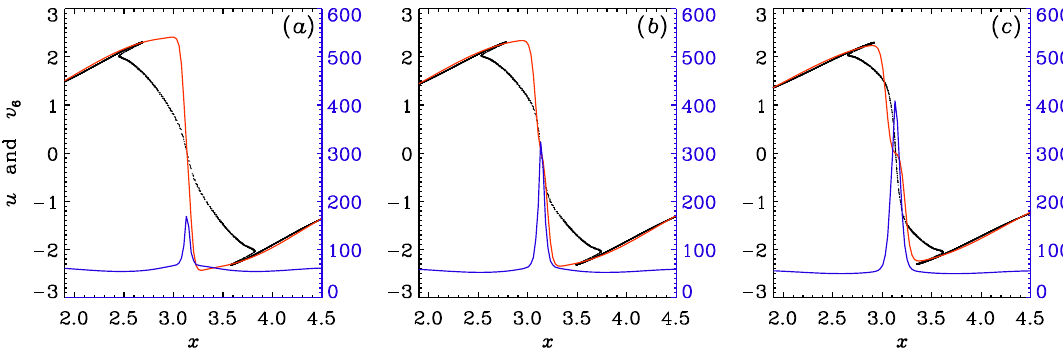}
\caption{
Velocity of particles with radius $a_6$ (black dots) and fluid velocity
(red lines), as well as the gas density (blue lines and axes on the right)
at $t=1.18$ (a), close to the time $t_*=1.16$ when the shocks
meet and the gas density develops a peak.
Panels (b) and (c) show the same at $t=t_*+\tau_6=1.22$ and
$t_*+2\tau_6=1.28$.
}\label{fig:pprelvel}
\end{figure}

The result of this experiment is shown in \Fig{fig:pbut}, where we plot
$xt$ diagrams of $\rho$ and $n_p$ for different particle sizes.
For small particles with small Stokes numbers, the particles follow
the gas.
This can be seen by the similarity between panels (a) and (b).
To estimate $\St_{\rm int}$ in this case, we used $\Delta u=2.6$ and
$\Delta x=0.08$, so $\tau_{\rm f}=0.08/2.6\approx0.03$.
For our largest particles with $a_{p}=0.1$, we find $\rho\approx300$;
see the blue lines in \Fig{fig:pprelvel}.
We again used $\rho_{p}=10^3$, which yields
$\rho_{p}/\rho=1000/300\approx3$.
Therefore, $\tau_{p}=\sqrt{\pi/8}\,(\rho_{p}/\rho)(a_{p}/\cs)\approx0.2$,
so we have $\St_{\rm int}\sim 7$.

For our largest particle size with $\St_{\rm int}\approx7$,
the two counter-streaming particle clouds associated with the two
opposing shocks tend to run through each other owing to their large inertia,
as we can see from \Figp{fig:pbut}{f}.
For the intermediate size, where $\St_{\rm int}\approx2$, however, a sizeable
particle cloud is left behind at the original position of the collision
of the two shocks; see \Figp{fig:pbut}{e}.
This critical value is close to unity, as one might have expected.
We conclude from this that a particle cluster will
form after the collision of two shocks if the Stokes number is
around unity.
Here, the Stokes number is based on the width and speed of the shock fronts.
This mechanism for particle clustering is fundamentally different from
the classical eddy mechanism of Maxey \& Riley and it operates only
for large enough Mach numbers.
For such flows, however, it may be the dominating mechanism.

The biggest uncertainty in our estimate of $\St_{\rm int}$ lies in the value of $\rho$.
In the following, we focus on the particles in bin~6,
and denote the corresponding velocity by $v_6$ and
the response time of those particles by $\tau_6$.
To determine the relevant value of $\rho$, we show in \Fig{fig:pprelvel} the
profiles of $\rho(x)$ together with those of $v_6(x)$ and $u(x)$ at times
$t_*+\tau_6/3$, $t_*+\tau_6$, and $t_*+2\tau_6$, where $t_*=1.16$ is the time
when the shocks meet and the gas density develops a peak.
We see that the peak in $\rho$ reaches values of around 400, but at the time
$t_*+\tau_6$, the particles will have slowed down considerably.
Therefore, the relevant density to be used is the temporally averaged
density at the peak until that time, which is below 300.

\subsection{Three-dimensional simulations}

In this section, we present our main result concerning the detection of
two separate clustering mechanisms through the Stokes number dependence
of the spectral particle number density.
Before presenting this, we discuss several peripheral aspects of the problem:
we first demonstrate that the effect of statistical noise on the power spectra
can be eliminated and we show how the results depend on the Reynolds number.
We also show that the results are insensitive to the choice of the drag
law (Stokesian versus Epstein drag).
The Eulerian approach is only used to determine its limits of
applicability at small Stokes numbers and the lack of agreement at
larger ones.
We finish with a demonstration of the artifacts caused by using a
shock viscosity, which is avoided in the bulk of this paper.




\begin{table}
    \centering
    \resizebox{\hsize}{!}{
    \begin{tabular}{ccllllccclccccc}
     Case    &forcing& $f_0$&$\delta t_{\rm f}$& $\nu$ & $\Ma$ & $\Reyn$ & $\Reyn_\lambda$ &
     $\tau_{\rm kin}$ & $\tau_{\rm Kol}$& $\St_{\rm int}^{\min}$ & $\St_{\rm int}^{\max}$ &
     $\St_{\rm Kol}^{\min}$ & $\St_{\rm Kol}^{\max}$ \\
\\
    \hline            
    V1  & vort.& 0.02 & --- & 0.001 & 0.15 &103 & 53&  4.3 & 1.8  & $7.3\times10^{-4}$ &  7.3 & $1.7\times10^{-3}$ &  17 & \\
    V1S & vort.& 0.02 & --- & 0.001 & 0.15 &103 & 53&  4.4 & 1.8  & $5.1\times10^{-4}$ &  5.1 & $1.2\times10^{-3}$ &  12 & \\    
    V2  & vort.& 0.2  & --- & 0.01~ & 0.67 & 45 & 32&  1.0 & 0.56 & $3.1\times10^{-3}$ &  31  & $5.5\times10^{-3}$ &  55 & \\
    V2a & vort.& 0.2  & --- & 0.02~ & 0.59 & 20 & 18&  1.13 & 0.81 & $2.8\times10^{-3}$ &  28 & $3.9\times10^{-3}$ &  39 & \\ %
    V2b & vort.& 0.2  & --- & 0.005 & 0.71 & 95 & 52&  0.94 & 0.39 & $3.3\times10^{-3}$ &  33 & $8.0\times10^{-3}$ &  80 & \\ %
    V3  & vort.& 0.5  & --- & 0.05  & 1.00 & 13 & 13 & 0.67 & 0.50 & $4.7\times10^{-3}$ &  47 & $6.3\times10^{-3}$ &  63 & \\ 
    V3a & vort.& 0.5  & --- & 0.02  & 1.14 & 38 & 26 & 0.59 & 0.32 & $5.3\times10^{-3}$ &  53 & $9.9\times10^{-3}$ &  99 & \\ 
    C1  & comp. & 0.5  & 0.5 & 0.005 & 0.19 & 25 & 19&  3.53& 1.89 & $8.9\times10^{-4}$ &  8.9 & $1.7\times10^{-3}$ &  17 & \\
    C1a & comp. & 0.5  & 0.5 & 0.002 & 0.19 & 64 & 29&  3.47 & 1.22 & $9.0\times10^{-4}$ &  9.0 & $2.6\times10^{-3}$ &  26 & \\
    C1.5& comp. & 1.5  & 1.0 & 0.015 & 0.39 & 17 & 15&  1.72 & 1.16 & $2.4\times10^{-2}$ &  54  & $3.6\times10^{-2}$ &  80 & \\
    C2  & comp. & 4.0  & 1.0 & 0.02~ & 0.76 & 25 & 13&  0.88& 0.57 & $3.6\times10^{-3}$ &  36  & $5.5\times10^{-3}$ &  55 & \\
    \end{tabular}
    }
    \caption{Summary of our simulations; ``comp'' and ``vort'' refer to
    compressive and vortical forcings, respectively.
    \newline
    \label{tab:models}}
\end{table}

\subsubsection{Overview of the different runs}

In the previous sections, we studied several aspects of particle clustering in 
idealised one-dimensional simulations.
We will now proceed by turning our attention to fully three-dimensional turbulence
simulations.
As described in \Sec{sec:model}, two kinds of forcings are employed in this work.
In \Tab{tab:models}, run names starting with ``V'' use vortical forcings,
while those starting with ``C'' adopt spherical expansion wave forcing
(compressive forcing).
The numbers behind those letters indicate different forcing strengths,
which yield different Mach numbers.
Different Reynolds numbers are indicated by letters a and b.
We also list the ranges $[\St_{\rm int}^{\min},\St_{\rm int}^{\max}]$
and $[\St_{\rm Kol}^{\min},\St_{\rm Kol}^{\max}]$ of Stokes numbers,
as defined in \Eq{eq:stokes}.
The run with Stokes drag is denoted by the letter S at the end.
We also compare with corresponding Eulerian models
(\Tab{tab:models_eulerian}), where we have included
an artificial viscosity and diffusivity needed to
stabilise the simulations; see \Eqs{momentump}{densityp}.
For simulations with larger Mach numbers, the mesh must be refined
in order to resolve the shocks.
This means that the mesh spacing is significantly smaller than the
Kolmogorov scale, and, hence, that the Reynolds number must be decreased
in order to confine the computational cost.

\begin{table}
    \centering
    \begin{tabular}{ccllllccllll}
     Case    &forcing& $f_0$&$\delta t_{\rm f}$& $\nu$ & $\nu_{p}$ & $\Ma$ & $\Reyn$ & $N_{\rm mesh}$ \\
    \hline            
    V1  & vort.& 0.02 & --- & 0.001 & 0.005 & 0.15 &103 & 256 \\
    V2  & vort.& 0.2  & --- & 0.01~ & 0.01~ & 0.67 & 45 & 512 \\
    C1  & exp. & 0.5  & 0.05& 0.002 & 0.001 & 0.19 & 96 & 512 \\
    C2  & exp. & 4.0  & 1.0 & 0.05~ & 0.05~ & 0.72 & 14 & 512 \\
    \end{tabular}
    \caption{Summary of Eulerian runs.
    For all these runs, the artificial diffusivity ($\kappa_p$) equals the artificial 
    viscosity ($\nu_p$).
    \newline
    \label{tab:models_eulerian}}
\end{table}

\begin{figure}
\centering
\includegraphics[width=.8\textwidth]{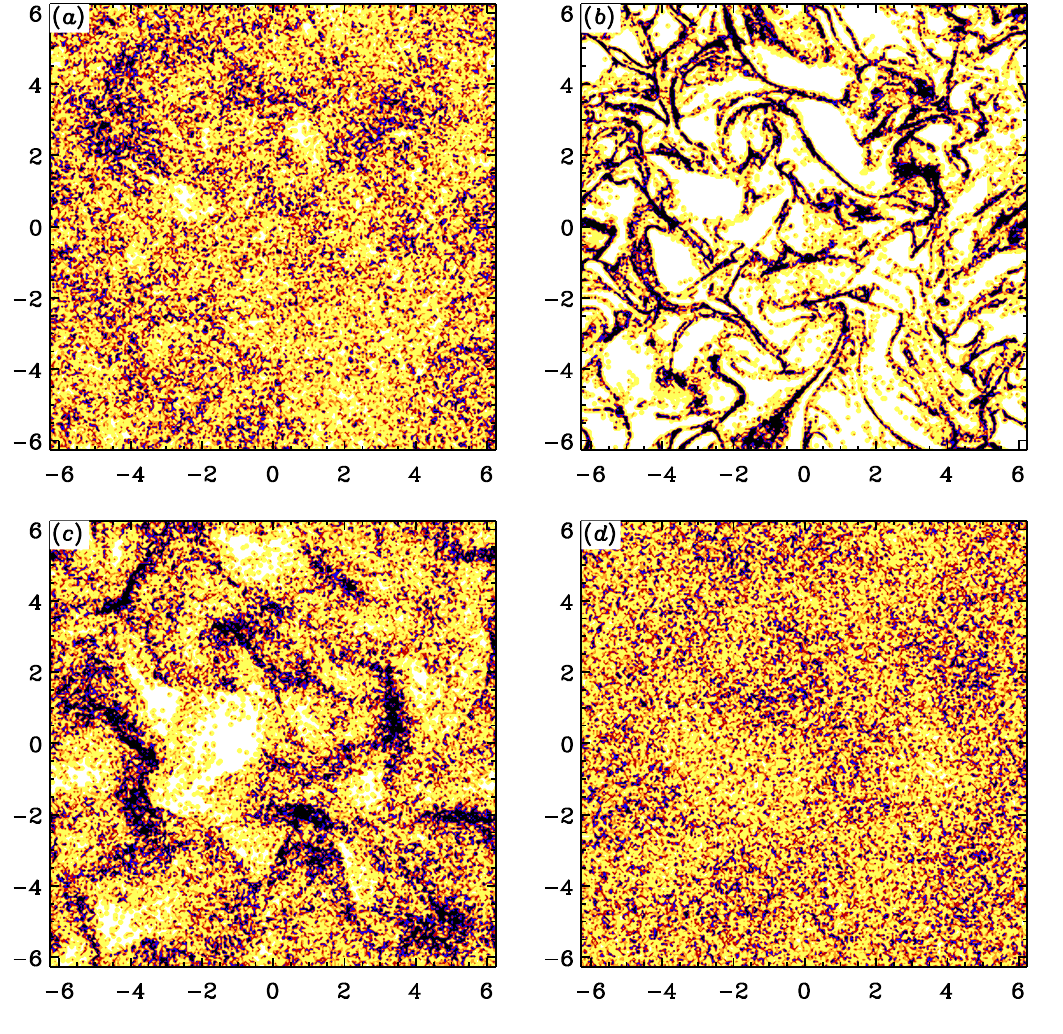}
\caption{
Contour plots of particle number density for (a) $\St_{\rm int}=$ 0.033,
(b) 0.33, (c) 3.3, and (d) 33 for case V2b. 
Dark shades denote high densities.
The particle number density has been integrated over the perpendicular
direction for four mesh zones.
\label{fig:pnp_ap}}
\end{figure}

\begin{figure}
\centering
\includegraphics[width=\textwidth]{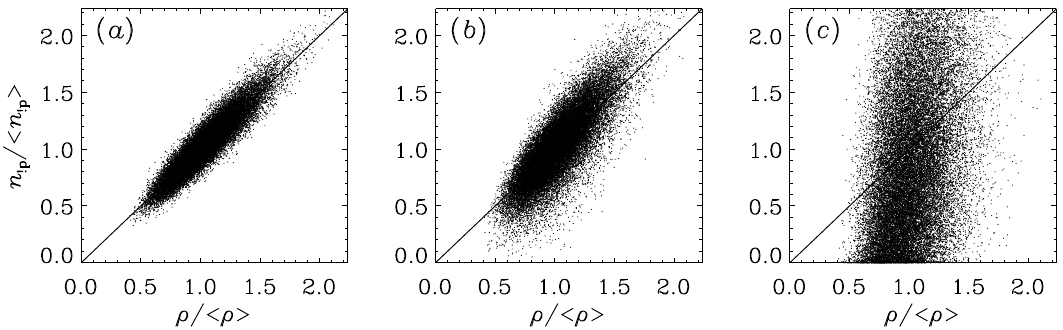}
\caption{Scatter plot of particle number density as a function of
fluid density for the three smallest particle sizes of case V2b. 
The solid line denotes the diagonal.
\label{fig:ppcoarse}}
\end{figure}

\begin{figure}
\centering
\includegraphics[width=.6\textwidth]{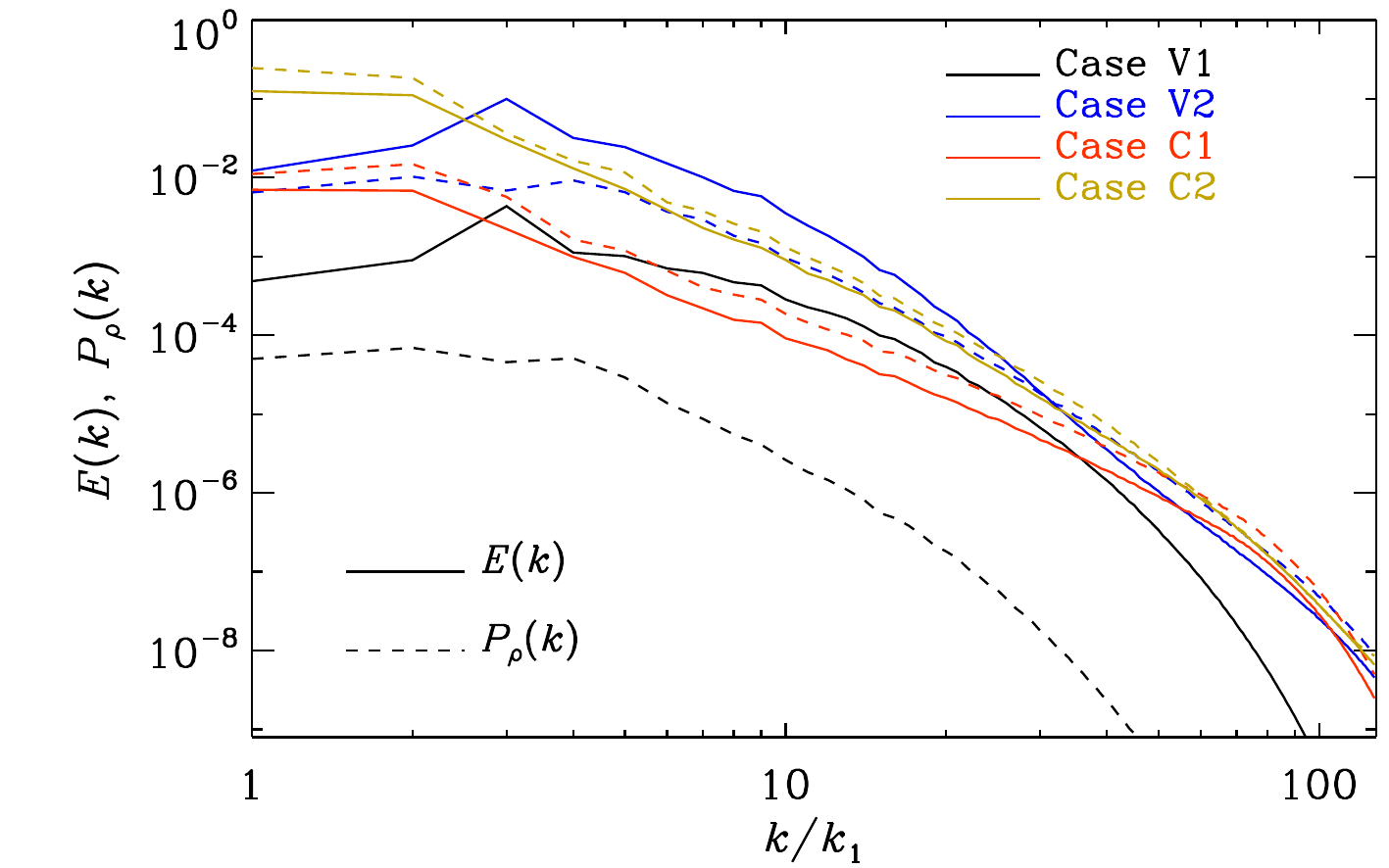}
\caption{$E(k)$ (solid lines) and $P_\rho(k)$ (dashed lines)
  for our main cases (V1, V2, C1, and C2).
\label{fig:fluid_spectra}}
\end{figure}

Contour plots of particle number density are shown in \Fig{fig:pnp_ap}
for Run~V2b.
It is clearly seen that the clustering is strongest for our intermediate
Stokes numbers, $\St_{\rm int}=0.33$ and $3.3$; see \Figsp{fig:pnp_ap}{b}{c}.
For the very smallest and largest Stokes numbers, we see almost no
clustering; see \Figsp{fig:pnp_ap}{a}{d}.
In those cases, the values of $\alpha$ are 0.7 and 7, respectively,
where we have used $L_{\rm f}=\kf^{-1}$.

In \Fig{fig:ppcoarse}, we show scatter plots of the particle number density as 
a function of fluid density for all fluid grid cells in 
the domain. In the left panel, 
showing results for the smallest Stokes numbers, we see that there is
a strong correlation between the two. This is because these small 
particles follow the fluid almost perfectly, which means that,
when the fluid is compressed (high fluid density), the particle field is 
also compressed (high particle number density).
When the Stokes number is increased, but is still rather small, we see
in \Figp{fig:ppcoarse}{b} that the two fields are only weakly correlated.
Finally, for those Stokes numbers where we see the strongest clustering
in \Figp{fig:pnp_ap}{b}, we show in \Figp{fig:ppcoarse}{c} that there
is no correlation between particle and fluid densities.
We can also see the effect of the inertial clustering itself in that there is a large
number of grid cells without any particles ($n_p=0$), while there is
also a significant number of grid cells containing many particles ($n_p > 2$),
which is not the case for the smaller Stokes numbers.

\subsubsection{Kinetic energy and density power spectra}
\label{SpectralAnalysis}

In this paper, we make extensive use of particle power spectra.
In this context, it is useful to show first the relevant spectra for the gas.
In \Fig{fig:fluid_spectra}, we show kinetic energy spectra together with
power spectra of the gas density for cases V1, V2, C1, and C2.
The peak of the kinetic energy spectrum occurs at $k/k_1=3$ for the cases
with vortical forcings (V1 and V2).
This wavenumber indicates where the main power
of the external forcing is found.
For the cases with compressive forcing, however, the peak
in the kinetic energy spectrum is found for $k/k_1$ between 1 and 2.
The density power spectra follow the kinetic energy spectra fairly well,
although there is a vertical shift.
For the compressive forcing, we are driving strong flow
divergencies, which results in $P_\rho(k)$ being large compared to
$E_{\rm K}(k)$.
Since the vortical forcing is divergence-free, the corresponding density variation is small, as
seen through the smaller values of $P_\rho(k)$ compared to $E_{\rm K}(k)$.
We can also see that the extent of the vertical shift is larger for
smaller Mach numbers.
This is because the fluid is less compressed for smaller Mach numbers.

\subsubsection{Initial and tracer particle spectra}
\label{InitialAndTracer}

In the following, we study power spectra of particle number densities for particles
that are embedded in a fluid where turbulence is generated either with
vortical or compressive forcing (Runs~V1--V3 and Runs~C1--C2,
respectively).
Since particles are tracked in a Lagrangian fashion,
it is convenient to allocate each Lagrangian
particle to the nearest Eulerian neighbours in each direction in the fluid mesh.
In this way, for every particle size, we generate
a variable on the fluid mesh that contains the number of particles
that reside within or in the neighbourhood of a given grid cell.
These variables can now be used to calculate the particle power 
spectra for each particle size.
The size of the neighbourhood of grid cells that will be influenced by
a given particle depends on the interpolation scheme used, which
in our case is a second order linear scheme \citep{2007Natur.448.1022J}.

Initially, the particles are randomly distributed
over the entire simulation box. This means that the initial power spectra
for the different particle sizes are just white noise, which corresponds to
a $k^2$ spectrum; see \Sec{PowerSpectra}.
If every particle is associated solely with the very nearest grid point
of the Eulerian mesh, the $k^2$ scaling would then be valid for the full
wavenumber range.
For the current work, however, the contribution from a particle is distributed
over several nearby grid points through a linear interpolation scheme.
This means that the $k^2$ scaling will not extend all
the way to the largest wavenumbers.
Instead, for $k>k_\ast$, the spectrum, hereafter $P_{{n},{\rm noise}}$,
becomes less steep and eventually reaches a maximum, before it goes
down towards the very end.

In analogy with \Eq{eq:np}, we now get
\begin{equation}
  \frac{1}{2} \langle n_p^2\rangle
  = \int_{k_1}^{{k}_{\rm max}} P_{{n},{\rm noise}}(k)\,\dd k
  = \int_{k_1}^{{k}_{\rm max}} A\tilde{k}^2\,\dd k
  \equiv \frac{A}{3}\tilde{k}_{\rm eff}^3,
\end{equation}
where $\tilde{k}=k/[1+(k/k_\ast)^3]$ has been substituted for $k$ in order 
to account for the departure from $k$
for $k\gtrsim k_\ast\equiv\kappa k_{\max}$, with $\kappa\approx0.789$ being
a parameter proportional to the position of the local maximum of $P_n(k)$.
This implies that $P_{{n},{\rm noise}}=A\tilde{k}^2$.
Defining therefore $\kappa=k_\ast/k_{\rm max}$, we find
$\tilde{k}_{\rm eff}^3=
(1+\kappa)(1-\kappa+\kappa^2)/\kappa^3k_{\rm max}^3\approx {(1.45\,k_{\max})^3}$.
By re-arranging the above equation, the constant $A$,
defined in \Sec{PowerSpectra}, is found to be
$A=3(\langle n_p^2\rangle/2)/\tilde{k}_{\rm eff}^3
\approx0.49\langle n_p^2\rangle/k_{\max}^3$.
Together with \Eq{eq:P_n} we then
obtain the initial power spectrum of the particles as
\begin{equation}
  \label{E_np}
   P_{{n},{\rm noise}}\approx
   0.49\,\frac{\langle n_p^2\rangle B^2}{k_{\rm max}^3}\tilde{k}^2,
\end{equation}
where the constant on the right-hand side is defined as
$B=\langle \rho\rangle/\langle n_p\rangle$ and is introduced
to compensate for the fact that the fluid and particle 
density fields do not have the same mean value.
This compensation is required in order to obtain \Eq{eq:P_model}.

\begin{figure}
\centering
\includegraphics[width=.49\textwidth]{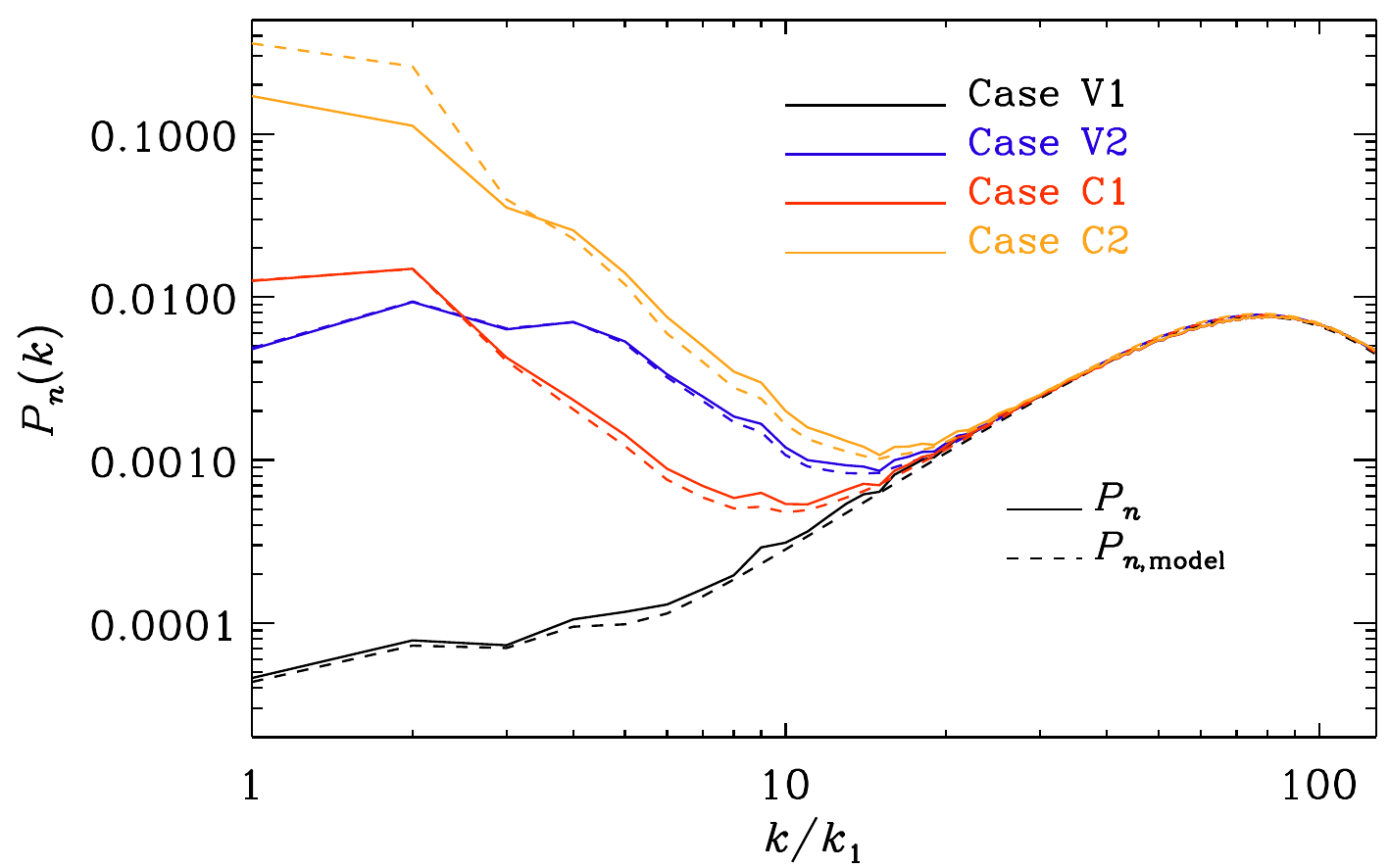}
\includegraphics[width=.49\textwidth]{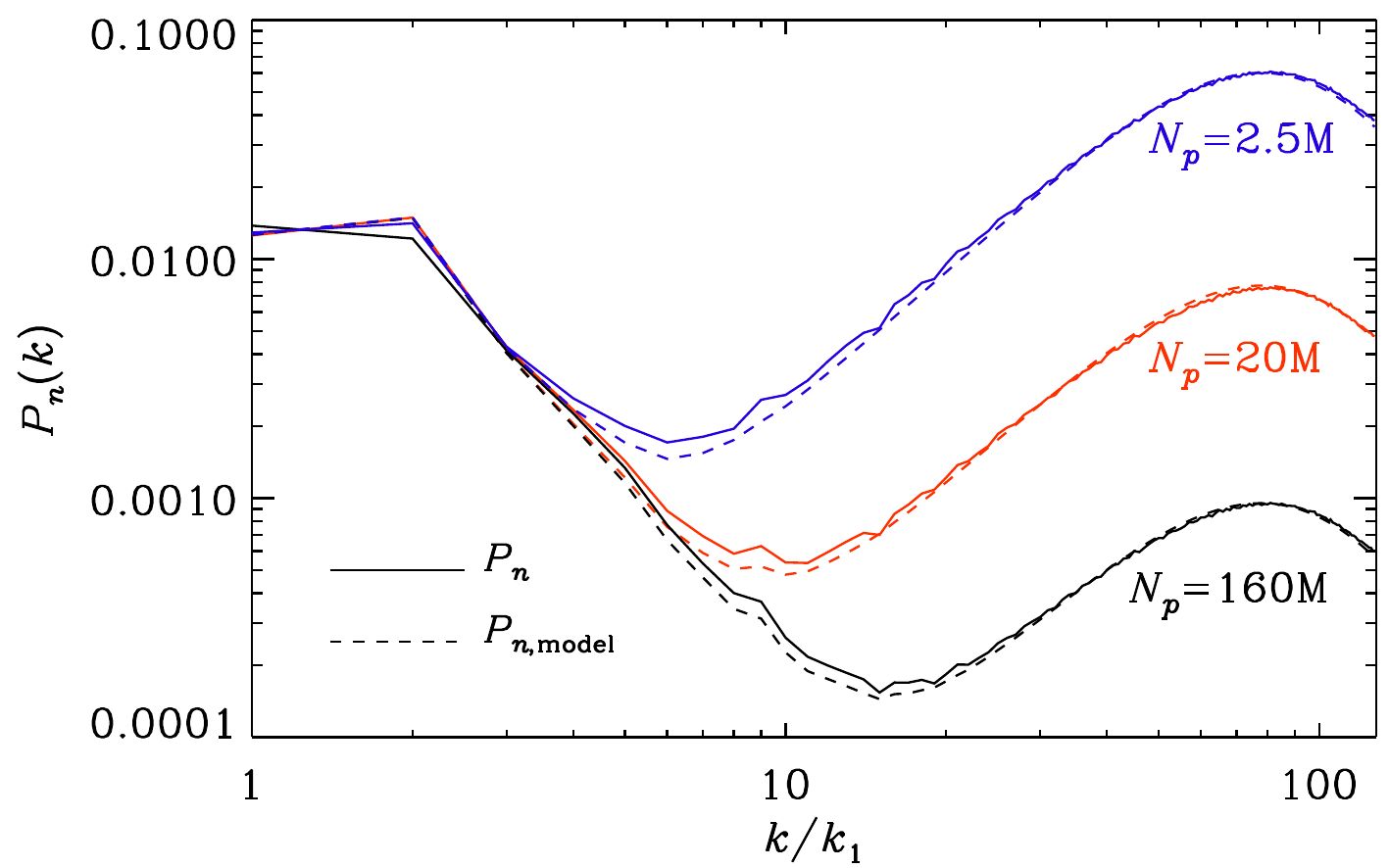}
\caption{
Left:
Power spectra of particle number
density for the smallest Stokes numbers, which are essentially tracer
particles (solid lines) for cases V1, V2, C1, and C2.
The dashed lines denote the model spectrum as presented in 
\Eq{eq:P_model}.
Right:
Comparison of numerical power spectra (solid) and model
spectra using \Eq{eq:P_model} (dashed) for the smallest particles of
Run~C1 with $N_p=2.5\times10^6$ (blue),
$20\times10^6$ (red), and $160\times10^6$ (black).
\label{fig:power_tracer}}
\end{figure}

Particles that are very small, having essentially vanishing Stokes
numbers, will follow the gas perfectly.
If the fluid is incompressible, particles will be re-shuffled owing
to turbulence, but their mean separation will be unchanged with time,
which means that there is no particle clustering.
Hence, the power spectrum of tracer particles in an incompressible
fluid would equal $P_{{n},{\rm noise}}$
for all times. If, however, the fluid is compressible,
the compression of the fluid may be so strong that the resulting fluctuations
in particle number density becomes larger than the white noise.
The particle power spectrum will then be the same as the one for the fluid density.
For tracer particles we therefore expect the power spectrum of the particles
to be given by
\begin{equation}
\label{eq:P_model}
  P_{n,{\rm model}}(k) = P_\rho(k) + P_{{n},{\rm noise}}(k).
\end{equation}
In \Fig{fig:power_tracer}({\it a}), we compare the calculated particle
spectra from the simulations $P_n(k)$ (solid lines) with the modelled
spectra $P_{n,{\rm model}}(k)$ (dashed lines) for the four runs V1, V2,
C1, and C2 for the lightest particles.
We see that the calculated and modelled spectra are remarkably similar
for all cases.
The $k^2$ part of the spectrum is a real physical effect and is a
result of having a finite number of particles in the simulation, such
a feature cannot be seen if the particles are tracked by the Eulerian
approach. Instead, one would then see that the particle spectra follow
the fluid density spectrum for all wavenumbers.
Unless the particle suspension consists of an infinite number of
particles, this behaviour is incorrect and is due to the fact that the
Eulerian approach treats the particle suspension as a continuous fluid
and not as a collection of a finite number of discrete particles.

In \Fig{fig:power_tracer}({\it b}), we show the numerical particle spectra for 
the particles with the smallest Stokes number of
Run~C1, obtained with three different numbers of particles (solid lines).
The numerical results are then compared with the model data given by
\Eq{eq:P_model}. We see that the model results reproduce the simulation 
results for all particle numbers.
But, more importantly, we note that the effect of the finite number of 
particles becomes less dominant when the number of particles is increased, 
which is as expected. 
For the black line ($N_p=160\times 10^6)$,
10 times more particles than fluid mesh points were used. 
This highlights the difficulty in exploring weak clustering at large wavenumbers 
for large Reynolds numbers.

\subsubsection{Comparison of cases with Epstein and Stokesian drag}

\begin{figure}
\centering
\includegraphics[width=.49\textwidth]{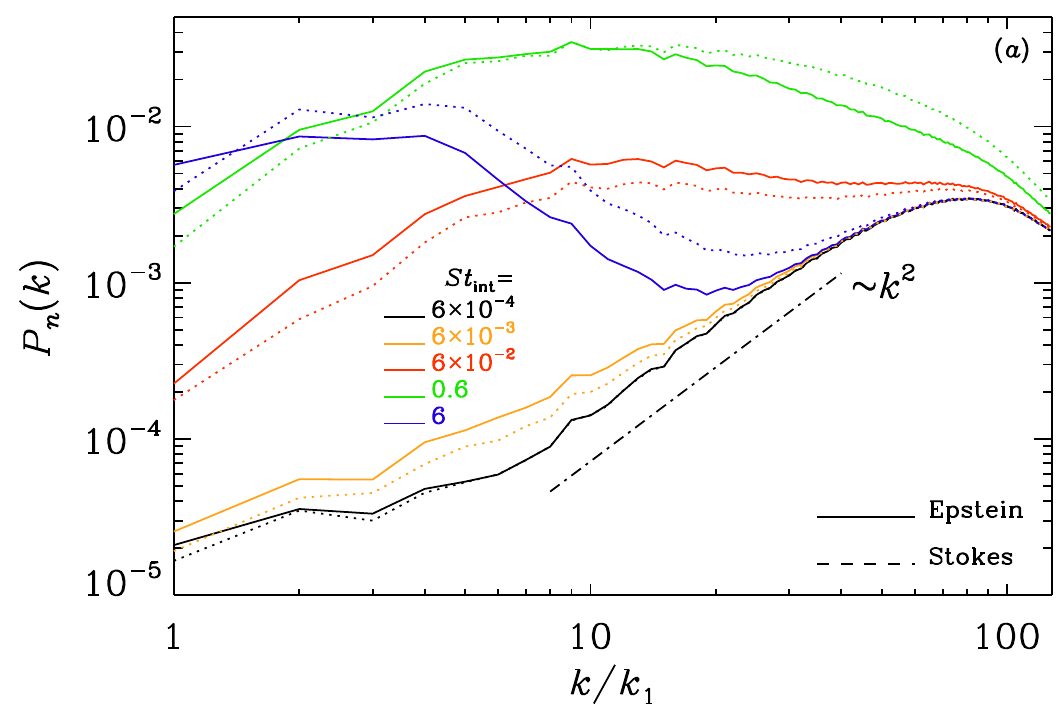}
\includegraphics[width=.49\textwidth]{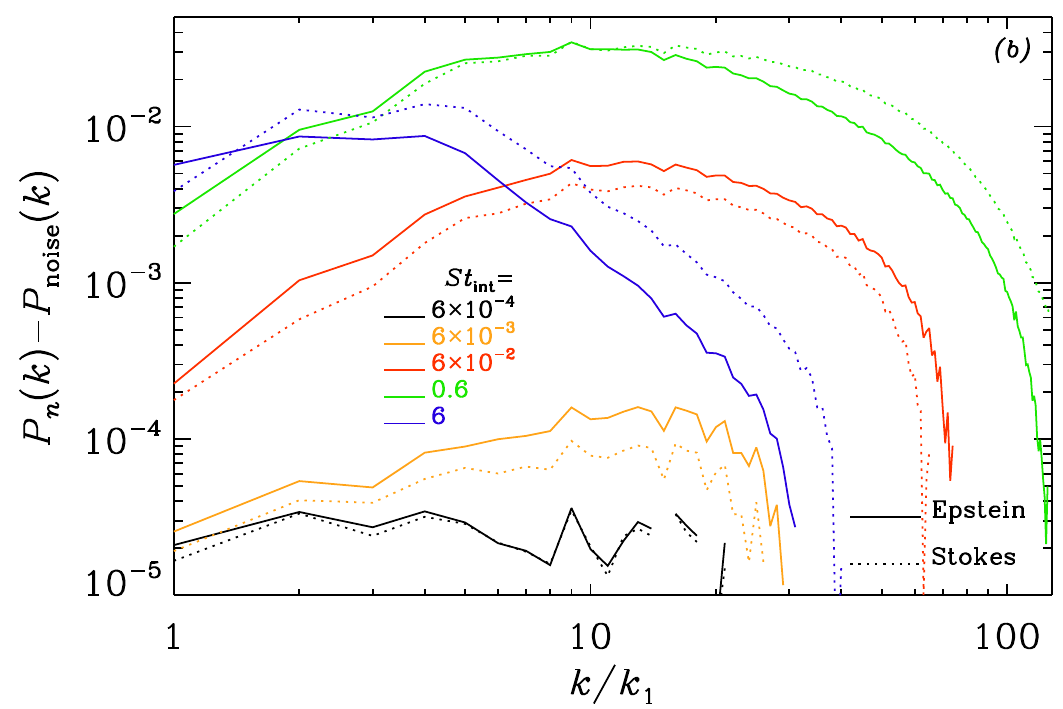}
\caption{
Comparison of power spectra of particle number densities 
for Run~V1 using Epstein (solid lines) and Stokes drag (dotted
lines) for $\St_{\rm int}=6\times10^{-4}$ (black), $6\times10^{-3}$ (orange),
$6\times10^{-2}$ (red), $0.6$ (green), and $6$
(blue) for (a) $P_n(k)$ and (b) $P_n(k)-P_{\rm noise}$.
\label{fig:Ep_vs_Stokes}}
\end{figure}

For rarefied gases, the molecular mean free path is large compared to the
particle size, and the response time is then given by the
Epstein time, as presented in Eq.~(\ref{tau_Ep}).
For a dense (continuous) fluid, however, where $\Kn<1$,
the response time is given by the modified
Stokes time; see Eq.~(\ref{tau_St}).
To compare the effect of using these two response times,
we show in \Fig{fig:Ep_vs_Stokes}
power spectra of particle number densities for simulations using
Epstein and Stokes drags for approximate Stokes numbers between
$6\times 10^{-4}$ and 6.
We see that the two reflect rather similar trends.
At least part of the remaining discrepancies can probably be 
explained by
small differences in the actual Stokes numbers for the two drag laws.
Both for small and large Stokes numbers, we see a $k^2$ spectrum,
indicative of random particle distributions at high wavenumbers 
(small scales).
Initially, particles of all sizes were randomly distributed,
and hence they had a $k^2$ spectrum.
However, the reason that we still see a $k^2$ spectrum at later
times are different for the smallest and largest particles.
The largest particles are so heavy that their response times are far
too long for the particles to be able to react to the turbulent eddies
associated with the smallest spatial and temporal scales.
This is why the power spectra of the heavier particles
still show their initial $k^2$ spectrum at small scales.
The smallest particles, by contrast, are being re-shuffled by the turbulence
and therefore maintain a random particle distribution.
They do not have enough inertia to move from one fluid element to another,
which is a requirement for the particles to form inertia-based clusters.
They are fundamentally different from the short-lived
increase in the particle number density, which occurs always when the fluid
volume in which the tracer particles reside, is compressed -- for
example due to the passing of an acoustic wave or shock.

\subsubsection{Applicability of the Eulerian approach for particles}
\label{Applicability_Eulerian}

\begin{figure}\begin{center}
\includegraphics[width=.49\columnwidth]{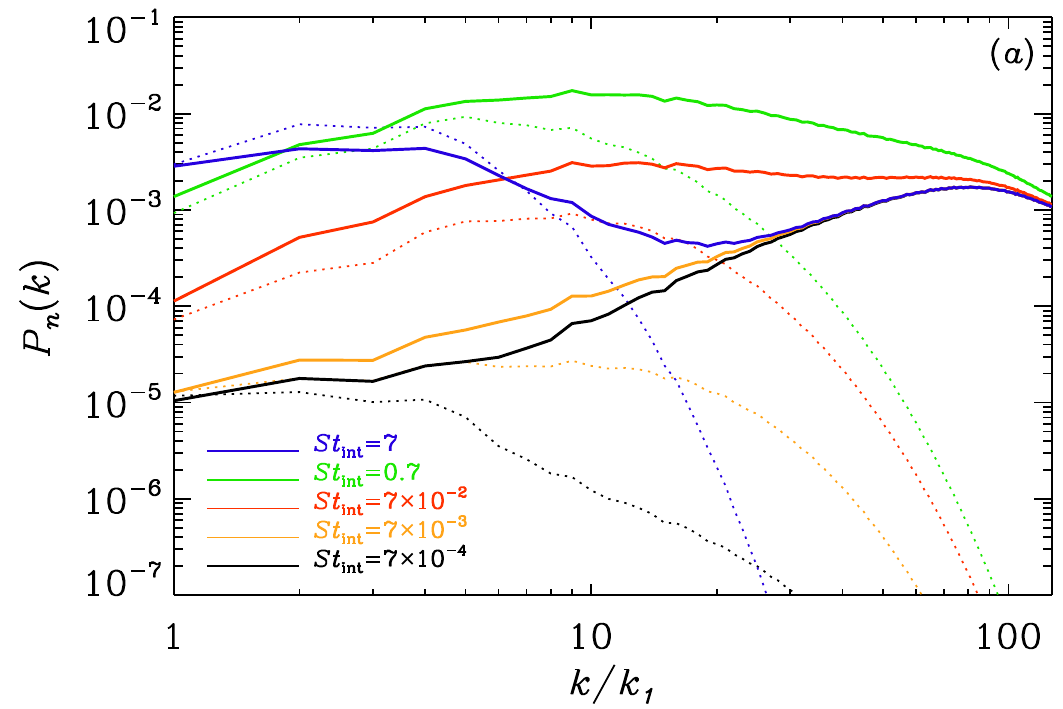}
\includegraphics[width=.49\columnwidth]{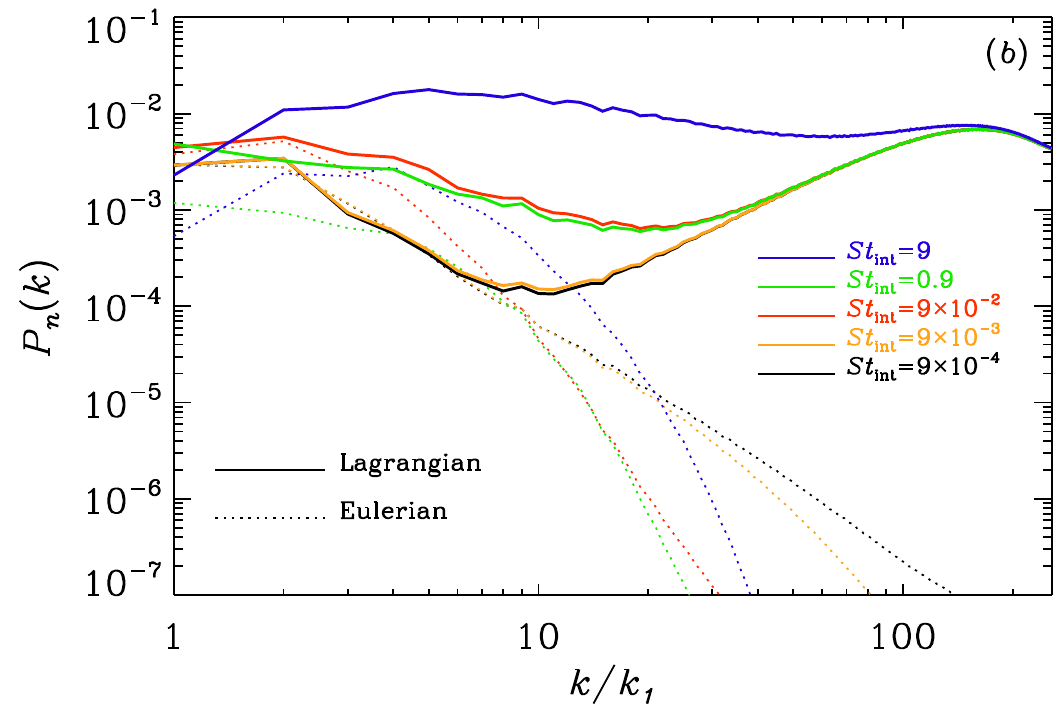}
\includegraphics[width=.49\columnwidth]{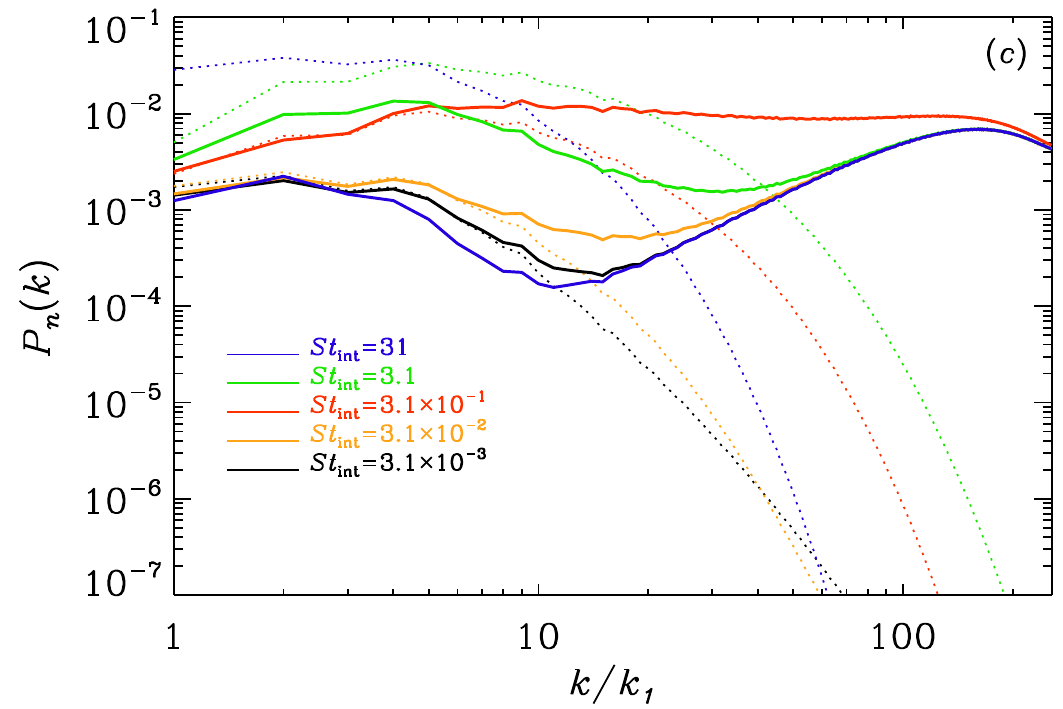}
\includegraphics[width=.49\columnwidth]{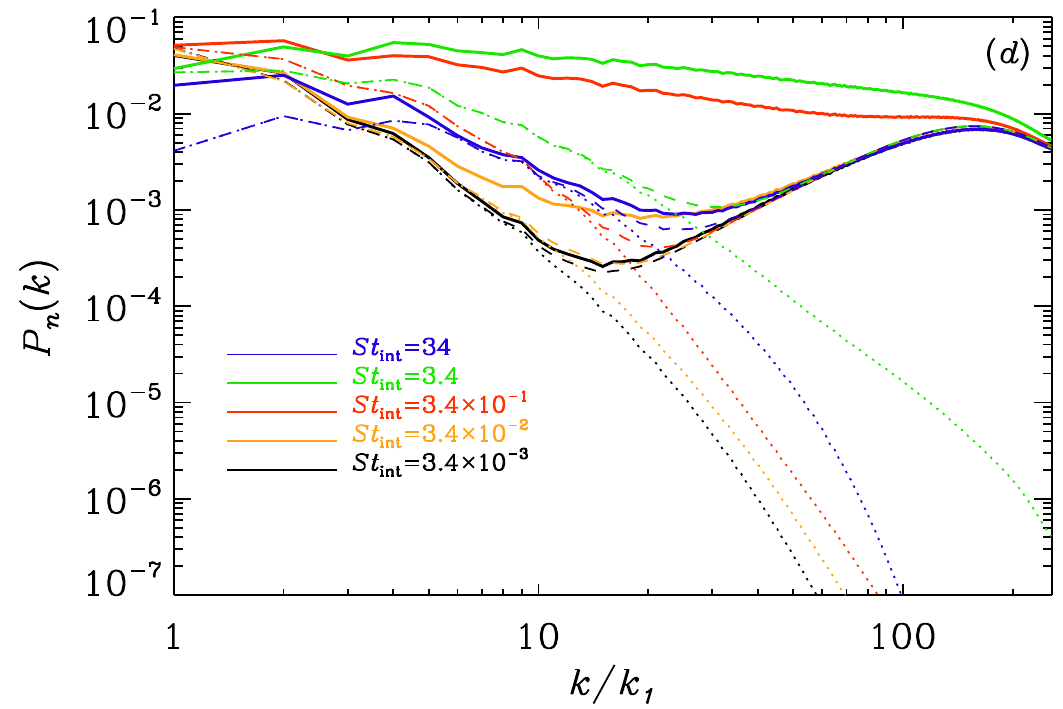}
\end{center}\caption[]{
Comparison of power spectra of particle number densities for
Runs~V1 ({\em a}), C1 ({\em b}), V2 ({\em c}), and C2 ({\em d}). 
The dashed lines in panel ({\em d}) are obtained by
adding $P_{{n},{\rm noise}}$, as given in \Eq{E_np}, to
the spectra obtained from the Eulerian simulation of Run~C2.
}\label{pcomp_spec_christer_helexp}\end{figure}

In \Fig{pcomp_spec_christer_helexp}, particle power spectra for
different Stokes numbers are shown for our four main runs, V1, V2, C1, and C2.
The solid lines correspond to results obtained with the Lagrangian approach
for the particles, while the Eulerian particle approach was used for
the simulations visualised by the dotted lines.
For V1 and V2, as shown in \Fig{pcomp_spec_christer_helexp}({\it a})
and ({\it c}), we see that the two approaches yield similar results for
the smaller Stokes number up to the wavenumbers where the $k^2$ scaling
commences in the Lagrangian simulations, which corresponds to $k \sim 4
k_1$ for the $5\times4$ million particles chosen for these simulations
with five different radii.
When the Stokes number is increased (different colours), the Lagrangian
and Eulerian spectra for $k/k_1 \leq 4$ are still comparable for V1,
but for V2 we see a clear difference for the largest Stokes numbers,
and not just for the largest wavenumbers; see also \cite{Boffetta+07}
and \cite{Shotorban+09} for similar results.

We recall that in the 1-D simulations, the critical value of
$\St_{\rm int}$ for the occurrence of caustics was around 0.3;
see the discussion at the end of \Sec{Applicability_Eulerian}.
In the present case of 3-D turbulence, we also see a difference
between the Eulerian and Lagrangian simulations near
$\St_{\rm int}=0.3$, but only for the spherical expansion waves;
see \Figp{pcomp_spec_christer_helexp}{d}.
When the turbulence is driven compressively, as in Runs~C1 and C2, which are
shown in \Figsp{pcomp_spec_christer_helexp}{b}{d}, we see that 
the Eulerian and Lagrangian approaches yield comparable results only for the very 
smallest Stokes numbers.
Thus, the range of applicability of the Eulerian approach depends on
the type of forcing and is more restrictive in the compressive case.
For the other cases, the differences are rather small, except perhaps
for the heaviest particles.
This difference could also be caused by our usage of artificial
viscosity and diffusivity in the Eulerian simulations.
At large wavenumbers, on the other hand, there is always a large
difference, but this is mainly caused by the effect of noise.

For Run~C2 in \Figp{pcomp_spec_christer_helexp}{d}, we also show 
the power spectra from the Eulerian approach with the contribution
from $P_{{n},{\rm noise}}$ of \Eq{E_np} being added.
This models a power spectrum that is accounting for the noise from a
finite number of Lagrangian particles; see the dashed lines.
For the smallest Stokes number, we see that the solid and dashed
lines almost overlap, but for all other Stokes numbers the resemblance 
is poor. This means that, except for the very smallest Stokes
numbers, the differences between the particle power spectra obtained
with the Lagrangian and Eulerian particle approaches are \emph{not}
primarily due to the noise contribution of the Lagrangian approach.
Furthermore, from the results shown in \Fig{pcomp_spec_christer_helexp},
we can also conclude that the Eulerian approach should \emph{not} be
used to track particles unless the Stokes number is low.

As expected from the discussion above (\Sec{InitialAndTracer}),
we notice a $k^2$ behaviour for the smallest Stokes numbers; see
\Fig{pcomp_spec_christer_helexp}.
This applies, of course, only to the Lagrangian simulations
with a finite number of particles, and cannot be seen in the
corresponding Eulerian simulations.
However, there is a clear departure from the $k^2$ behaviour as the Stokes
number is increased.
The reason is that particle inertia now starts to have an effect
on the clustering. This clustering is not due to fluid compression,
but rather due to other inertia-based clustering 
mechanisms, such as the Maxey--Riley mechanism.

For intermediate Stokes numbers, the particle power spectra resemble
those presented in \cite{2018JFM...836..932H}.
In that paper, the authors speculate that the peak in the particle
power spectra is associated with the similarity in the characteristic
time scales of the turbulence and the response time of the particles.
However, such a connection could not be confirmed in their work owing
to their limited Reynolds number.
We see here the same trend as found by Haugen et al., namely that
the individual maxima of the spectra are insensitive to the Stokes
number.
We expect this to change at higher Reynolds numbers and higher resolution.

\subsubsection{Reynolds number dependence}

\begin{figure}\begin{center}
\includegraphics[width=.49\columnwidth]{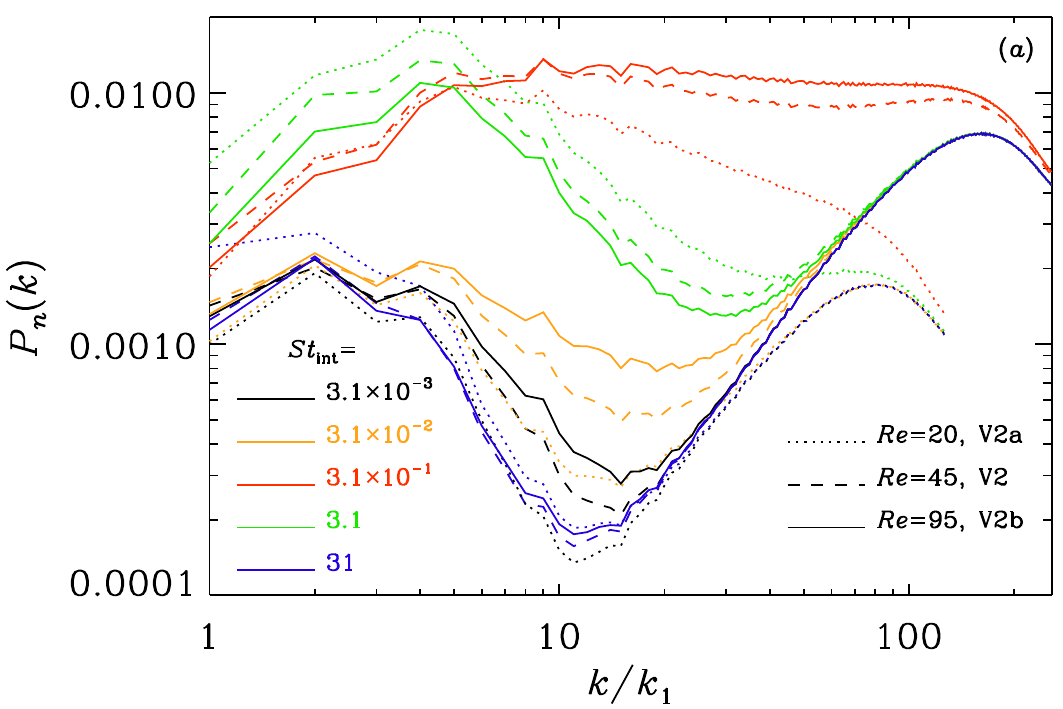}
\includegraphics[width=.49\columnwidth]{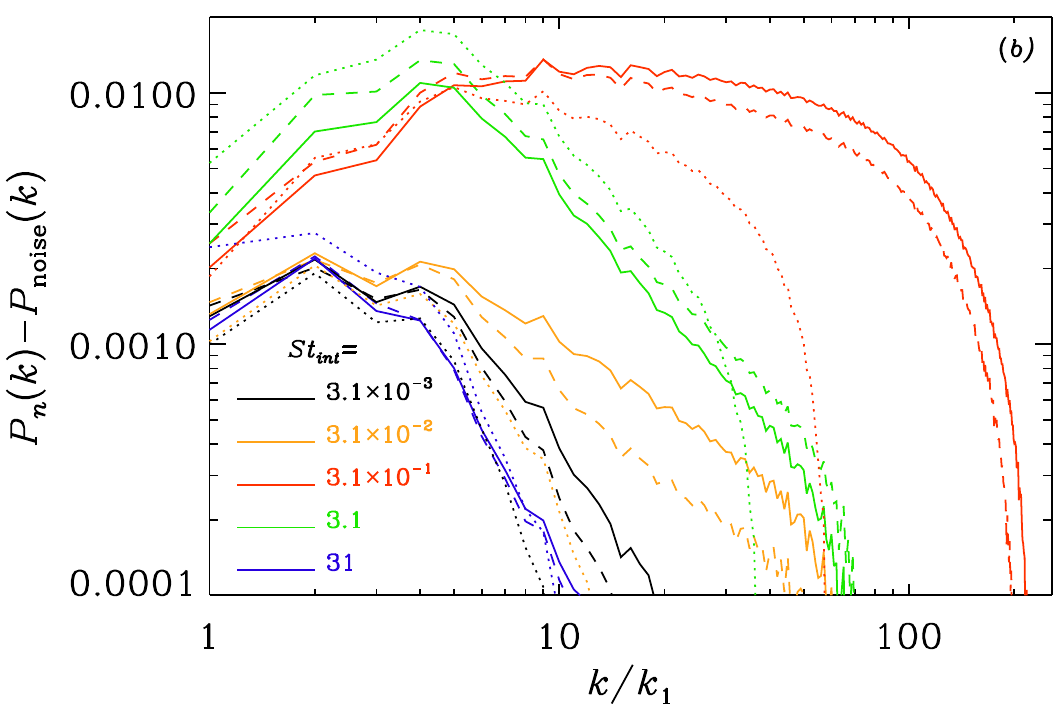}
\end{center}\caption[]{
    Comparison of power spectra of particle number densities 
    for Run~V2 (dashed lines) with Run~V2a (dotted lines) and
    Run~V2b (solid lines) for (a) $P_n(k)$ and (b) $P_n(k)-P_{\rm noise}(k)$.
}\label{pcomp_spec_christer_Re_hel}\end{figure}

In order to investigate the nature of the inertia-based clustering
further, we would like to run simulations with much larger Reynolds numbers.
In the DNS, this becomes very costly when the Mach number is large and
shocks need to be resolved.
In \Fig{pcomp_spec_christer_Re_hel}, we show the power spectra for
different Stokes and Reynolds numbers.
For the smaller Stokes numbers, an increase in $\Reyn$ leads to an
increase in spectral power.
For the larger Stokes numbers, the trend is opposite.
Looking at the Kolmogorov-based Stokes number ($\St_{\rm Kol}$), however, we see that,
for a given value of $\Reyn$, we get more power and hence more clustering
when the Kolmogorov-based Stokes number is closer to unity.
For the Reynolds numbers obtained here, it therefore seems that it is the 
Kolmogorov based Stokes number that controls the strength of the
clustering, not the one based on the integral scale ($\St_{\rm int}$).
The same has also been found in other low Reynolds number studies \citep{Bec07b,Baker2017}.
As the Reynolds number is increased, however, one eventually
reaches a point where particles with $\St_{\rm int}$ around unity will be much slower than
the smallest turbulent eddies and they will therefore be totally decoupled from 
the Kolmogorov scale. Hence, the clustering cannot be determined by
$\St_{\rm Kol}$ in such cases.
The nature of this large-scale particle clustering still remains to
be understood, because much larger resolution would be needed.

\begin{figure}\begin{center}
\includegraphics[width=.99\columnwidth]{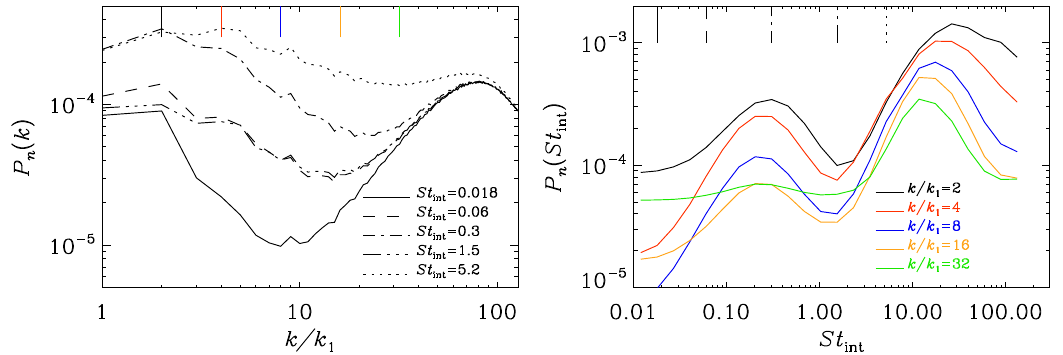}
\end{center}\caption[]{
$P_n(k)$ and $P_n(St_{\rm int})$ for
Run~C1 with more particle sizes.
The different line types in the left panel, marked in the legend, correspond
to the line types of the short vertical lines on the upper abscissa of the right panel.
Likewise, the different colours in the right panel, indicated in the legend, correspond
to the colours of the short vertical lines in the left panel.
}\label{non-mono-spec}\end{figure}

\subsubsection{Clustering mechanisms}

From Run~C1, shown in \Figp{pcomp_spec_christer_helexp}{b}, we see that
the spectra for $\St_{\rm int}=0.09$ and 0.9 are very similar.
This may be an indication of a non-monotonic behaviour of the
spectral evolution with Stokes number.
In order to investigate this further, we perform a new simulation
that is identical to Run~C1 except that we now include many more
closely spaced particle sizes (Stokes numbers).
The spectra for some of these Stokes numbers are shown in the left-hand
panel of \Fig{non-mono-spec}.
From this we see that the spectral power increases from $\St_{\rm int}=0.018$
up until $\St_{\rm int}=0.3$, before it decreases again as we move towards $\St_{\rm int}=1.5$.
Finally, a clear increase is seen for larger Stokes numbers.
In the right-hand panel of \Fig{non-mono-spec} we plot the power for
five different wavenumbers as a function of the particle Stokes number
in order to see this non-monotonic behaviour more clearly.
Here we see that, for all wavenumbers shown, 
the power spectra attain two distinct
maxima: the first is around $\St_{\rm int}=0.3$, while the other is 
found between $\St_{\rm int}=10$ and 30.
It is not immediately clear what is causing this non-monotonic behaviour,
but we argue here that it is due to a change in the relative importance
between two different particle clustering mechanisms.
Since compressive forcing was used for this run, the two competing
mechanisms are most likely (i) the shock-clustering mechanism, as described
in Section~\ref{sec:1D_B}, and (ii) the classical Maxey--Riley
clustering mechanism.

\begin{figure}\begin{center}
\includegraphics[width=.99\columnwidth]{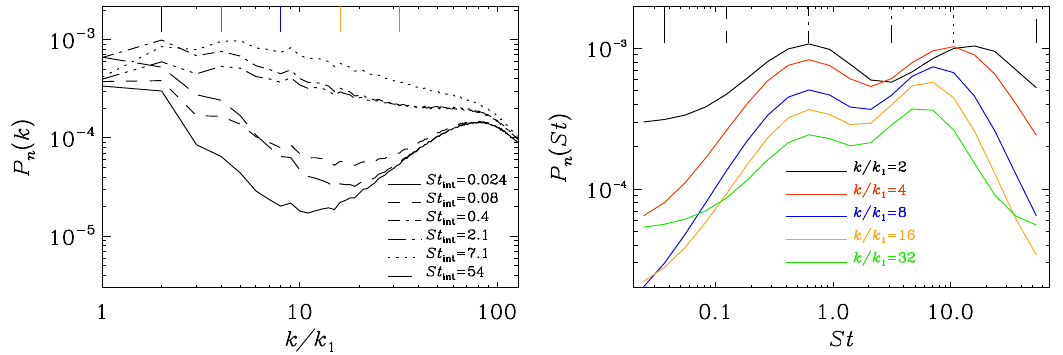}
\end{center}\caption[]{
Similar to \Fig{non-mono-spec}, but for
Run~C1.5 with more particle sizes.
}\label{non-mono-spec_E1p5}\end{figure}

\begin{figure}\begin{center}
\includegraphics[width=.99\columnwidth]{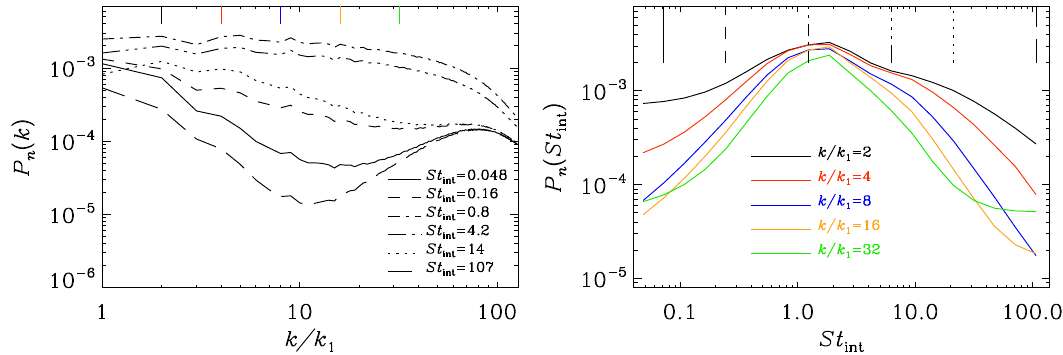}
\end{center}\caption[]{
Similar to \Fig{non-mono-spec}, but for
Run~C2 with more particle sizes.
}\label{non-mono-spec_E2}\end{figure}

If one of the peaks is due to the shock-clustering mechanism, we would
expect it to be stronger as the Mach number is increased.
We therefore perform an intermediate simulation (Run~C1.5) where we
increase the Mach number to $\Ma=0.39$ to investigate this.
The results are shown in \Fig{non-mono-spec_E1p5}, where it is clearly
seen that the first peak has become substantially stronger, and also moved
somewhat to the right.
The second peak is, however, almost unchanged, except for a smaller
shift to the left.
This seems to indicate that it is the first peak that is due to the
shock-clustering mechanism.
There may well be parallels with the simulations of \cite{Yang+14} and \cite{Zhang+16},
which used, however, a shock-capturing scheme
and were therefore not DNS.
We now proceed by increasing the Mach number even further (Run~C2) and show the
results in \Fig{non-mono-spec_E2}. The first peak has now become so strong that
the second peak is only visible as a weak shoulder for $\St_{\rm int}\approx10$.
We will now continue by investigating the mechanism behind 
the second peak.

For the classical eddy mechanism of Maxey \& Riley, we expect the
clustering to depend primarily on the vortical part of the velocity field.
Hence, the Stokes number for the second peak should be
of the order of unity if the fluid time scale is calculated
based on the vortical part of the velocity field.
To check this, we have performed a Helmholtz decomposition of
the velocity field by computing the vector and scalar potential of
$\uu=\nab\times\psi+\nab\phi$.
The rms values of the corresponding velocity fields are given in
\Tab{tab:Helmholtz}.
We also list the estimated peak Stokes numbers from the simulations
at $k/k_1=32$ and those based on the vortical velocity field.
We see that the vortical peak Stokes numbers are 1.2 and 1.6 for Runs~C1
and C1.5, while for Run~C2, we only see an indication of a shoulder at
$\St_{\rm peak2}^{\rm vort}=1.6$, at least for smaller values of $k$.
This could be taken as an indication that the second peak is indeed
due to the classical eddy mechanism of Maxey \& Riley, or the
non-local clustering mechanism discussed by \cite{Bragg+15}.

\begin{table}
    \centering
    \begin{tabular}{llccccccc}
Run & $\urms$ & $\urms^{\rm vort}$ & $\urms^{\rm pot}$
& $\St_{\rm peak1}$ & $\St_{\rm peak2}$
& $\St_{\rm peak1}^{\rm vort}$ & $\St_{\rm peak2}^{\rm vort}$ \\
    \hline            
C1  &  0.170 & 0.017 & 0.169 & 0.15& 12 & 0.02 & 1.2 \\
C1.5&  0.352 & 0.11  & 0.334 & 0.6 &  5 & 0.19 & 1.6 \\
C2  &  0.865 & 0.200 & 0.842 &  2  & (7)& 0.46 &(1.6)\\
    \end{tabular}
    \caption{
Rms velocities for the Helmholtz decomposed velocities
together with the estimated peak Stokes numbers from the simulations
and those based on the vortical velocity field.
The numbers in parentheses are more uncertain. The reason for this is that,
for Run~C2, the second peak appears as a shoulder only, and no clear maximum can be identified.
}
    \label{tab:Helmholtz}
\end{table}

\begin{figure}
\centering
\includegraphics[width=\textwidth]{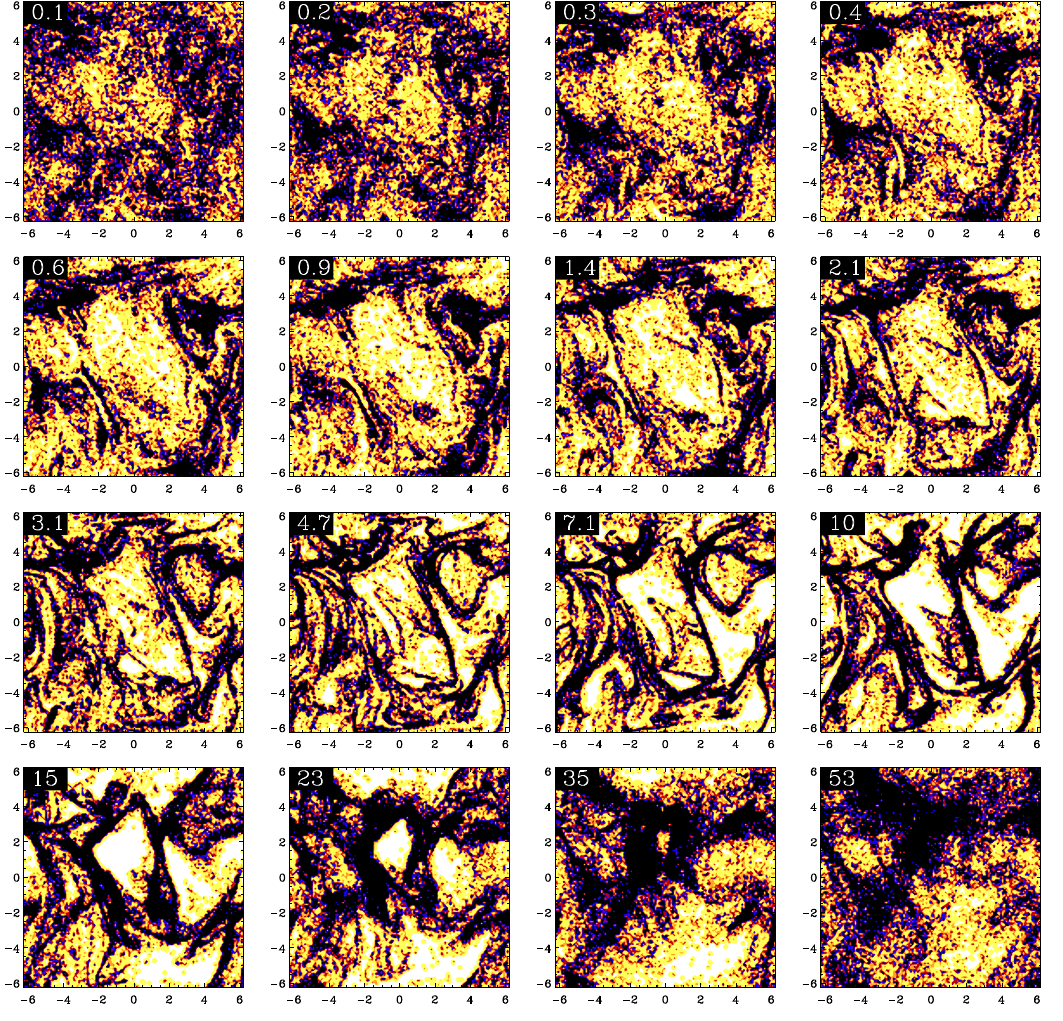}
\caption{
Similar to \Fig{fig:pnp_ap}, but for Run~C1.5 showing
contour plots of particle number density for $\St_{\rm int}$
in the range from 0.1 to 53.
\label{fig:pnp_ap_E1p5}}
\end{figure}

In \Fig{fig:pnp_ap_E1p5}, we show the particle concentrations
for Run~C1.5 in order to determine if we can see any trace of the two
different mechanisms behind the particle clustering.
We see that clustering is now apparent for a very broad range of Stokes
numbers, ranging from $\St_{\rm int}=0.3$ to $50$.
Both for small and large values of $\St_{\rm int}$ do we see blob-like clusters,
while for intermediate values the structures are more sheet-like.
Other than that, there is no real difference in the morphology of 
structures between small and large Stokes numbers.

\begin{figure}\begin{center}
\includegraphics[width=.99\columnwidth]{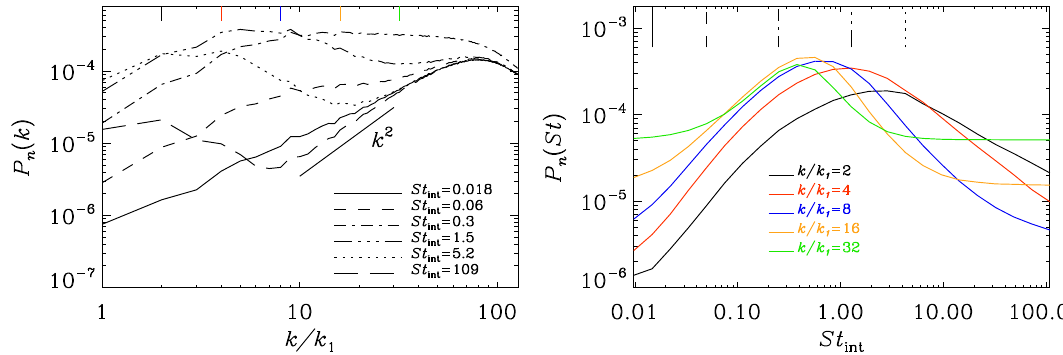}
\end{center}\caption[]{
Similar to \Fig{non-mono-spec}, but for Run~V1 with more particle sizes.
}\label{non-mono-spec_V1}\end{figure}

In \Fig{non-mono-spec_V1} we present similar results as in
\Figs{non-mono-spec}{non-mono-spec_E2}, but now for the low Mach-number
case with vortical forcing (Run~V1). 
Our results are therefore more similar to earlier ones for incompressible
turbulence \citep[see, for example,][]{Ireland+16a,Ireland+16b}.
Here there is no indication of anything more than a single peak.
This peak, which is due to the classical eddy-clustering of Maxey \& Riley, 
is found to be around $\St_{\rm int}\approx1$, as expected.

\subsubsection{RDFs}

\begin{figure}\begin{center}
\includegraphics[width=.99\columnwidth]{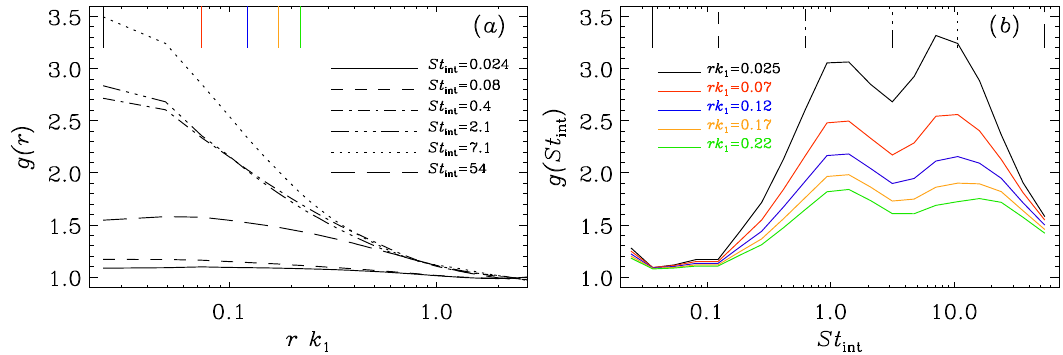}
\end{center}\caption[]{
RDFs for Run~C1.5, (a) shown as a function of $r$ for different Stokes
numbers, and (b) as a function of $\St$ for five different separations
($r k_1=0.025$, 0.07, 0.12, 0.17, and 0.22).
The different line types in panel (a), marked in the legend, correspond
to the line types of the short vertical lines on the upper abscissa of panel (b).
Likewise, the different colours in panel (b), indicated in the legend, correspond
to the colours of the short vertical lines in panel (a).
}\label{pc_comp_rrdf}\end{figure}

Typically, the RDF is mainly used for small distances of a few Kolmogorov
lengths, but here we are interested in larger distances, too.
To speed up the calculation in that case, we use the numbers
of particles at each mesh point and sum up the products of
particle numbers between all pairs of mesh points where the particle
number is finite.
We show in \Figs{pc_comp_rrdf}{pc_comp_rrdf_V1} RDFs for Runs~C1.5 and V1.
The abscissa is normalised by the smallest wavenumber in the domain,
$k_1=2\pi/L$.
This means that particle separations up to half the domain size are shown.
Owing to the discrete spacing of mesh points within the various shells,
the resulting $g(r)$ was not smooth, but this problem is readily
alleviated by normalising instead with an empirically determined
discrete version of $g(r)$ for a random particle distribution.
We note that RDFs based on the mean particle number per mesh point
can also be used for the Eulerian approach, in spite of its other
shortcomings.

In incompressible Kolmogorov-type turbulence, $g(r)$ tends to show
a gradual decline with increasing $r$ \citep{Salazar2008}.
This overall trend is also seen in the present cases of
compressible turbulence.
This is characteristic of the fractal nature of the particle
distribution.
In the case of Run~C1.5, however, we also see characteristic
peaks of $g(\St_{\rm int})$ for $\St_{\rm int}\approx1$ and 10.
These values of $\St_{\rm int}$ agree with those where enhanced clustering was
found in \Fig{non-mono-spec_E1p5}.
For the run with vortical forcing, we only see a single peak
both in the spectra and the RDFs as a function of Stokes number;
see \Figs{non-mono-spec_V1}{pc_comp_rrdf_V1}.

\begin{figure}\begin{center}
\includegraphics[width=.99\columnwidth]{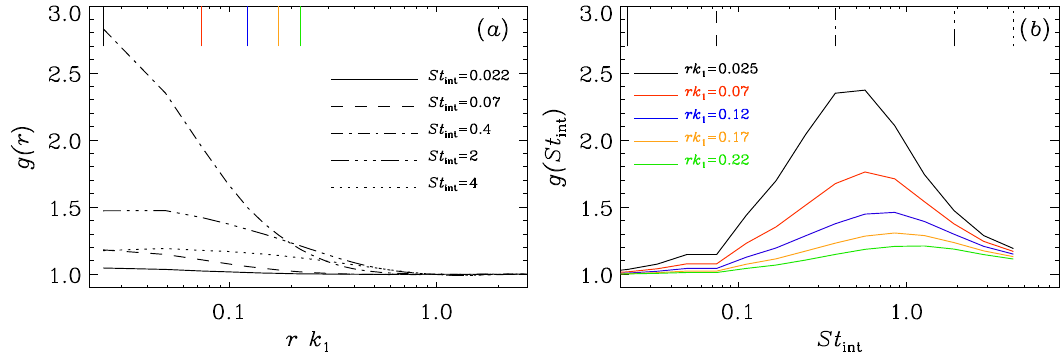}
\end{center}\caption[]{
Similar to \Fig{pc_comp_rrdf}, but for Run~V1.
}\label{pc_comp_rrdf_V1}\end{figure}

It may be useful to compare our results of large-scale clustering with
earlier ones by \cite{Saw+12}, who also claimed to have found large-scale
clustering in wind tunnel experiments.
In their case, however, such clustering was believed to be mainly the
result of their initially inhomogeneous field of particles.
They computed RDFs, which showed a characteristic shoulder at about
a hundred Kolmogorov scales.
Our simulations do not show such a shoulder, but this could be owing
to a lack of scale separation.
It would have been interesting to see their RDFs also as a function of
Stokes number in addition to just the separation.
In particular, it would then be important to include also larger values
of the Stokes number.

\begin{figure}\begin{center}
\includegraphics[width=.49\columnwidth]{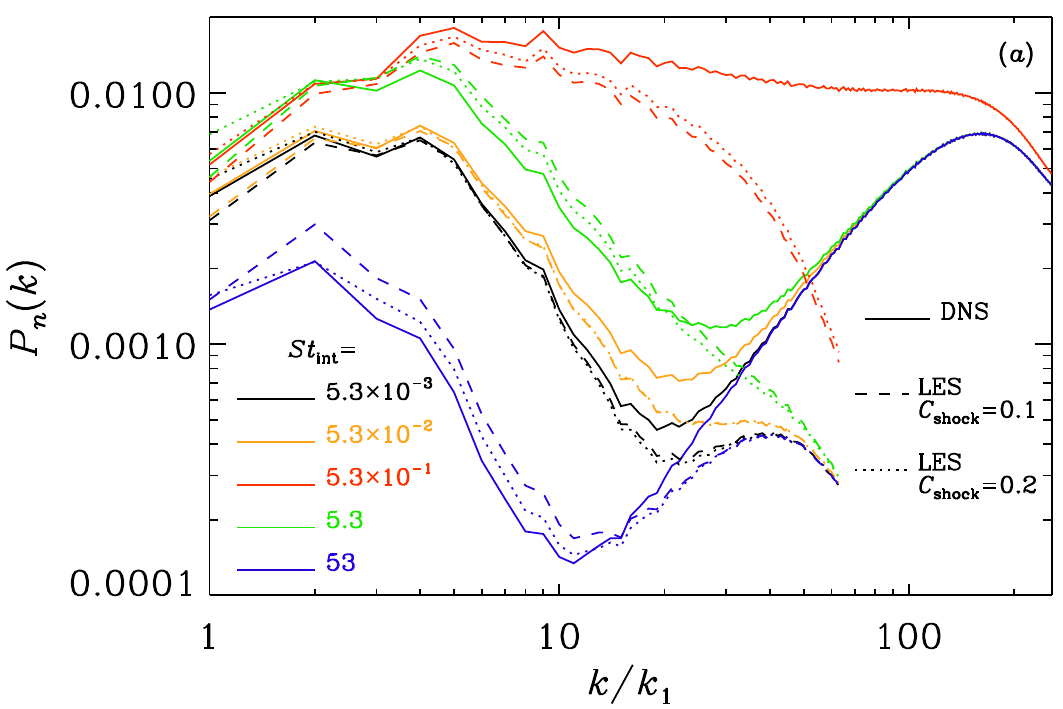}
\includegraphics[width=.49\columnwidth]{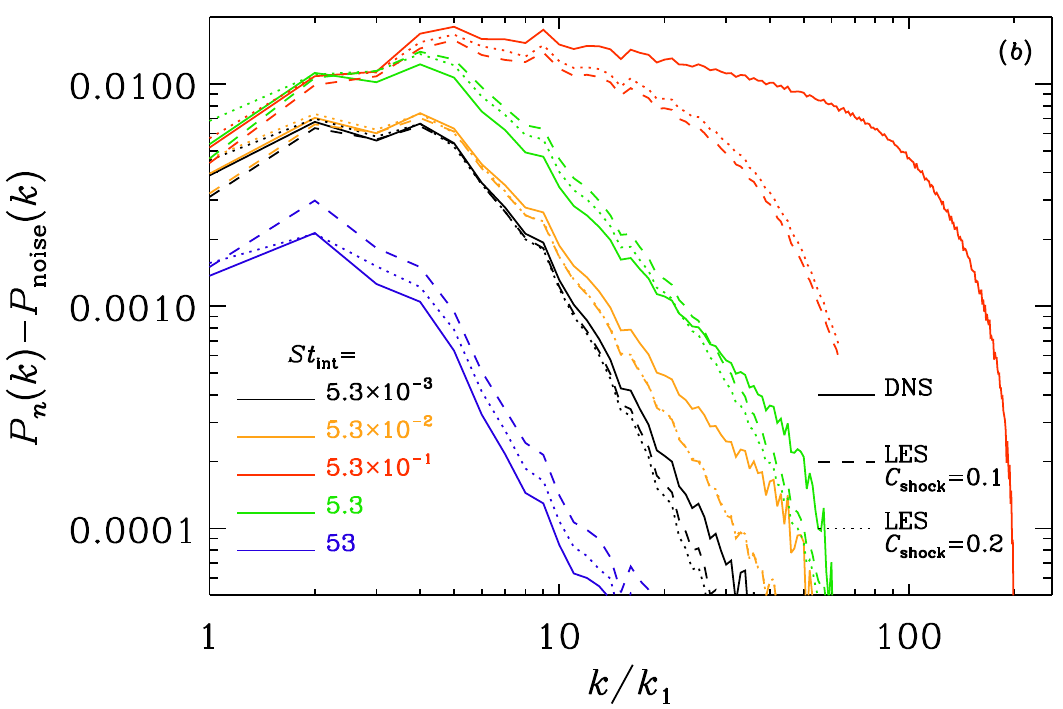}
\end{center}\caption[]{
Similar to \Fig{pcomp_spec_christer_Re_hel}, showing a
    comparison of Run~V3a (solid lines),
    as well as Run~V3s01 (dashed-dotted lines) and Run~V3s02 (dotted lines)
    for (a) $P_n(k)$ and (b) $P_n(k)-P_{\rm noise}$.
}\label{pcomp_spec_christer_Ma_hel}\end{figure}

\subsubsection{Shock capturing viscosity}

\begin{figure}\begin{center}
\includegraphics[width=\columnwidth]{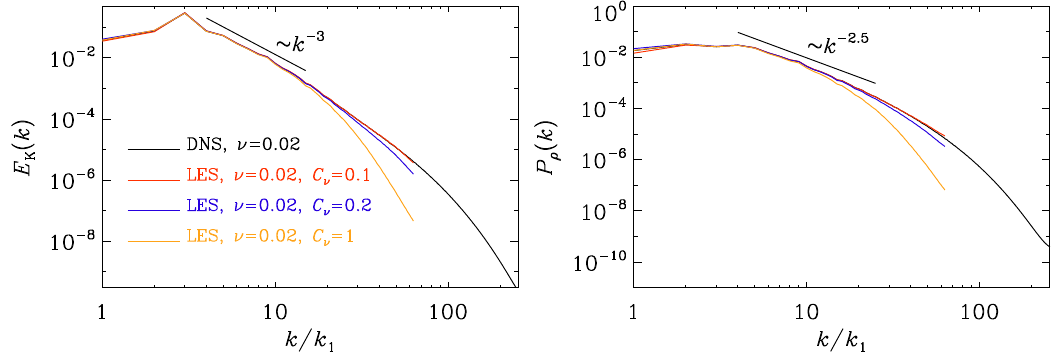}
\end{center}\caption[]{
$E_{\rm K}(k)$ (left) and $P_\rho(k)$ (right) for Runs~V3 and V3a
(DNS with $\nu=0.05$ and $0.02$, respectively) as well as Runs~V3s01,
V3s02, and V3s1 (LES with $C_\nu=0.1$, $0.2$, and, $1$ respectively).
}\label{pspec_kine_comp}\end{figure}

Finally, let us discuss how well the clustering results can be modelled
using lower numerical resolution together with a shock viscosity to
stabilise the code.
Such a shock viscosity was used in the LES of \cite{Yang+14} and
\cite{Zhang+16}, but it remained unclear to what extent this affected
the accuracy of their results.
To perform a meaningful comparison between DNS and LES,
it is interesting to have an even larger value of $\Ma$.
To be able to do this, it is interesting to have an even larger value of $\Ma$.
Therefore, we consider a DNS with $\Ma=1.14$ (Run~V3a) and compare with
two LES with different values of $C_{\rm shock}$; see \Eq{zeta_shock}.
The result for the density spectra is shown in \Fig{pcomp_spec_christer_Ma_hel}.
We see that for all values of $\St_{\rm int}$, except for $\St_{\rm int}=0.53$, 
the LES spectra (dashed and dashed-dotted lines) are close to the DNS
for $k/k_1<20$.
For $\St_{\rm int}=0.53$ (red lines), however, the agreement exists only up to
$k/k_1\approx6$.
Surprisingly, a similar departure is not seen in the kinetic energy and
fluid density spectra shown in \Fig{pspec_kine_comp}.
A major difference is, of course, that the LES do not resolve the
small length scales at all, which is also why their spectra are shorter.
The discrepancy in the particle density spectra between LES and DNS
therefore suggests that the clustering, which occurs mostly at those
intermediate values of $\St_{\rm int}$, depends on physical effects at
the scale of the shocks, corresponding to high wavenumbers.
If this is indeed the case, this departure between DNS and LES may
become worse at larger values of $\Ma$.
It will be interesting to revisit this question in future simulations
at higher resolution.

We should point out that the name LES is, in the present context,
somewhat of a misnomer, because here all the eddies are actually
resolved.
This is because the Kolmogorov scale in Run~V3 is about
three times as large as the mesh spacing; see \Tab{tab:eta}.
In the LES, the mesh spacing is four times larger than in the DNS,
which means that we are resolving down to almost the Kolmogorov scale.



\begin{table}
    \centering
    \begin{tabular}{lccccr}
     Run & $\ell_{\rm Kol}$ & $\delta x$ & $\epsK$ & $\Ma$ & $\Reyn$ \\
    \hline            
    V1  & 0.042 & 0.049 & 0.0003& 0.15 &101 \\
    V2b & 0.044 & 0.025 & 0.034 & 0.72 & 95 \\
    V3a & 0.080 & 0.025 & 0.199 & 1.14 & 38 \\
  V3s01 & 0.074 & 0.098 & 0.267 & 1.15 & 38 \\
  V3s02 & 0.076 & 0.098 & 0.234 & 1.14 & 38 \\
  C2    & 0.213 & 0.049 & 0.064 & 0.74 & 15 \\
  C2a   & 0.107 & 0.025 & 0.061 & 0.76 & 38 \\
  C2s05 & 0.106 & 0.098 & 0.065 & 0.75 & 37 \\
  C2s1  & 0.106 & 0.098 & 0.063 & 0.74 & 37 \\
    \end{tabular}
    \caption{
Summary of the Kolmogorov scale $\ell_{\rm Kol}=(\nu^3/\epsK)^{1/4}$,
mesh spacing $\delta x$, energy dissipation rate $\epsK$, as well as Mach
and Reynolds numbers.
    \newline
    \label{tab:eta}}
\end{table}

\begin{figure}\begin{center}
\includegraphics[width=.49\columnwidth]{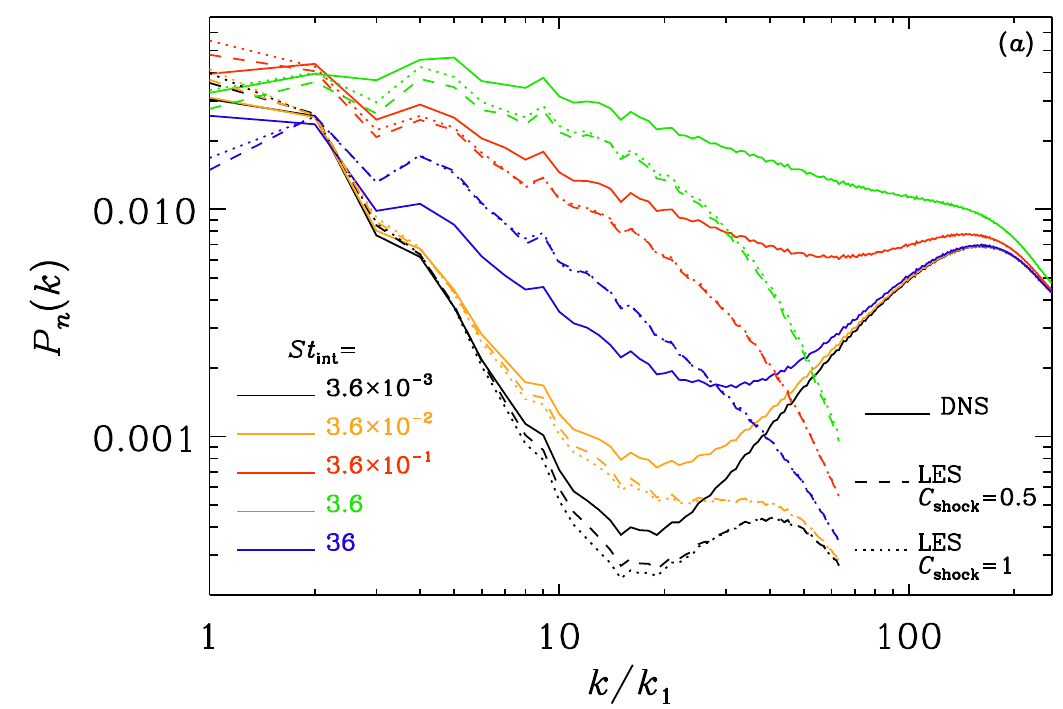}
\includegraphics[width=.49\columnwidth]{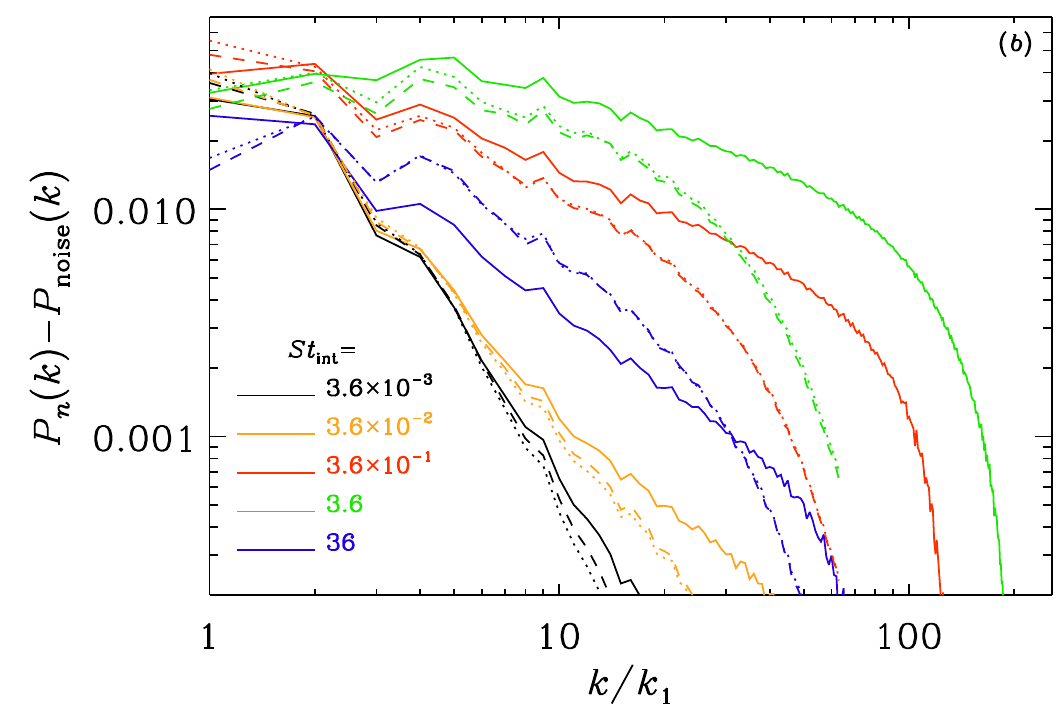}
\end{center}\caption[]{
    Comparison of Run~C2 (solid lines),
    as well as Run~C2s05 (dashed-dotted lines) and Run~C2s1 (dotted lines)
    for (a) $P_n(k)$ and (b) $P_n(k)-P_{\rm noise}$.
}\label{pcomp_spec_christer_Ma_hel_C2}\end{figure}

The quality of LES is worse for runs with compressive forcing; see
\Fig{pcomp_spec_christer_Ma_hel_C2}, where we show the results for Run~C2
with $\Ma=0.76$.
We see that for $\St_{\rm int}=3.6$ (green lines), the agreement between
DNS and LES is rather poor even for small values of $k/k_1$.
For $\St_{\rm int}=0.36$ (red lines), the agreement is slightly better,
but again only for small values of $k/k_1$.

\section{Conclusions}

In this work we have used particle power spectra to investigate
particle clustering for compressible isotropic turbulence. 
We have shown that, by plotting the dependence on the Stokes number
for a particular wavenumber, they are a particularly suitable tool for
identifying large-scale clustering owing to various clustering mechanisms
such as Maxey--Riley and shock clustering.
For studying small-scale clustering, the conventional RDFs
remain a more suitable tool.

We have studied the effect of using either the Epstein drag, which applies
for $\Kn\gg1$, or the modified Stokesian drag, which applies for $\Kn\ll1$.
As long as the particle radii are non-dimensionalised
with the Stokes number, the power spectra resulting from the two drag
laws turned out to be similar.
This supports the general usefulness of the Stokes number -- even
for compressible flows and very diverse drag laws.

When using the concept of power spectra to analyse particle clustering
of Lagrangian particles, it is important to realise that the number of
particles used will have an effect on the power at large wavenumbers
(small scales).
This is due to the fact that, if too few particles are used, there are
not enough particles to populate such small clusters and it becomes
impossible to identify clusters at small scales.
This effect is clearly seen through the presence of a $k^2$ contribution
in the power spectra at small scales.
The magnitude of the $k^2$ contribution is proportional to the inverse
of the total number of particles in the simulation.
If the Eulerian approach is used for the particles, it is implicitly
assumed that an infinite number of particles are involved.
This implies that the $k^2$ contribution to the power spectra is absent.
However, this only applies to cases with an infinite number of particles.

When using the Eulerian particle approach, multi-valued particle
velocities are not possible.
This is because the Eulerian approach cannot represent caustics, which
implies that only very small Stokes numbers would be modelled correctly.
Furthermore, for larger Stokes numbers, artificial diffusion and viscosity
must be used for the particle fluid in order to stabilise the simulations.
This can yield nonphysical results.
Hence, the Eulerian particle approach can only be used to simulate a
very large number of particles that are small enough so that they behave
almost as tracers.

\begin{figure}
\centering
\includegraphics[width=.8\textwidth]{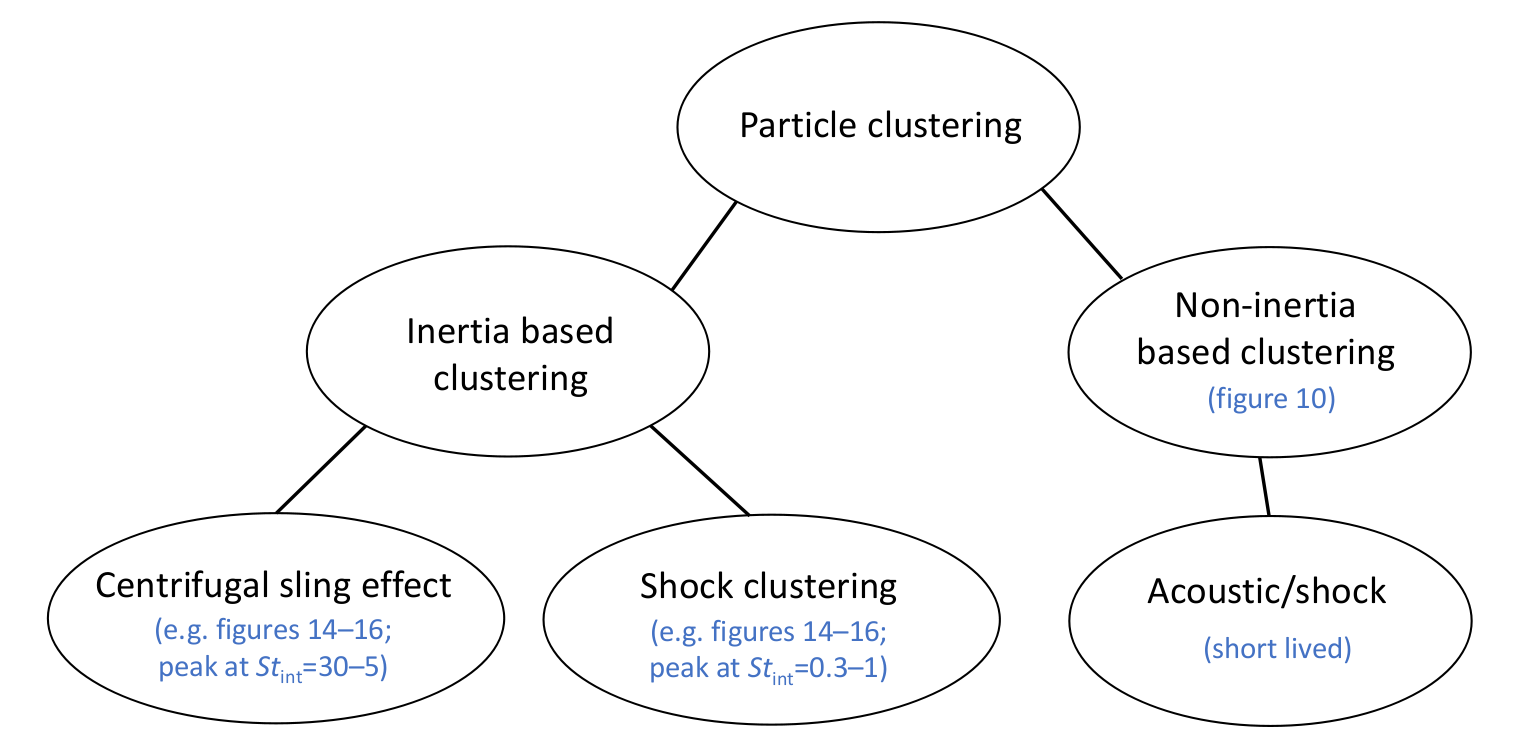}
\caption{
Sketch summarising the various clustering mechanisms discussed in this paper.
\label{fig:clustering}}
\end{figure}

The main finding of the present study is that there is a significant
difference in the clustering, depending on how the turbulence is generated.
For vortical forcing, the clustering peaks at
an integral scale-based Stokes number of around unity.
As already explained in \Sec{NonDimensionalNumbers},
there is an ambiguity regarding the most meaningful normalisation
of $\St_{\rm int}$, and one could argue for one that would make
its value smaller by a factor of $2\pi$.
However, when we use compressive forcing, we drive strong flow divergences.
In that case, we find that clustering peaks at
two different integral scale-based Stokes numbers, 
one somewhat below unity, and the other
at much larger Stokes numbers.
We argue that the first peak is explained by shock clustering,
similar to what was found by \cite{Yang+14} and \cite{Zhang+16},
while the second is the usual Maxey-Riley clustering
(based on the centrifugal sling effect), 
and its integral scale-based
Stokes number is found to be around unity if it is evaluated
based on the vortical part of the velocity field; see
\Fig{fig:clustering} for a summary.

In order to resolve shocks in high Mach number DNS, a very fine mesh
is required.
In many cases, interesting physics is, however, not
related to the internal structure of the shocks themselves,
but to the flows outside the shocks.
It is therefore often regarded as useful to model shocks through a shock
viscosity instead of resolving them, such that a coarser mesh can be used.
We investigated the effect on the particle clustering by using a 
shock-capturing viscosity to broaden the shocks.
For the simulations performed here, the shock-broadening
meant that the mesh was allowed to have four times less grid points in
each direction.
We found that the cases with shock-capturing viscosity reproduced the
results of a fully resolved DNS for the first decade of wavenumbers
rather accurately.
However, the relatively strong clustering at small scales that was found
for the DNS of particles with integral scale-based
Stokes numbers slightly less than
unity were not reproduced.
In view of these caveats, it would be interesting to revisit
the earlier work of \cite{Yang+14} and \cite{Zhang+16}.

\section*{Acknowledgements}
This work was supported by the Knut and Alice Wallenberg Foundation
through the grant Dnr.\ KAW 2014.0048 on
``Bottlenecks for particle growth in turbulent aerosols.''
We thank the Swedish Research Council, grant numbers 2015-04505 (LM)
and 2019-04234 (AB), for partial financial support.
The simulations were performed using resources provided by
the Swedish National Infrastructure for Computing (SNIC)
at the Royal Institute of Technology in Stockholm.
\vspace{2mm}

\noindent
{\em Code and data availability.} The source code used for the
simulations of this study, the {\sc Pencil Code} \citep{2020_JOSS},
is freely available on \url{http://github.com/pencil-code/}.
The DOI of the code is http://doi.org/10.5281/zenodo.2315093.
The simulation setup and the corresponding data are freely available on
\url{http://doi.org/10.5281/zenodo.4733175}; see also
\url{http://www.nordita.org/~brandenb/projects/isoth_expwave/} for easier access.

\vspace{2mm}
\noindent
{\bf Declaration of Interests.} The authors report no conflict of interest.

\appendix

\section{Forcing algorithms}
\label{ForcingAlgorithms}

The purpose of this appendix is to summarise the two types of forcings.
For vortical forcing, we choose
\begin{equation}
\ff(\xx,t)={\rm Re}\{{\cal N}\tilde{\ff}(\kk,t)\exp[i\kk\cdot\xx+i\varphi]\},
\end{equation}
where we select randomly at each time step a phase $-\pi<\varphi\le\pi$ and
the components of the wavevector $\kk$ from a discrete set of wavevectors
with average wavenumber $\kf$.
Here, $\xx$ is the position vector and
${\cal N}=f_0(\cs\kf\delta t)^{1/2}$
is a normalisation factor, where $\delta t$ is the time step and $f_0$
is an amplitude factor.
To ensure that $\tilde{\ff}$ is solenoidal, i.e., perpendicular
to $\kk$, we write is as
\begin{equation}
\tilde{\ff}({\kk})=(\kk\times\eee)/[\kk^2-(\kk\cdot\eee)^2]^{1/2},
\end{equation}
where $\eee$ is an arbitrary unit vector that is not aligned with $\kk$.

For compressive forcing with $\ff=-\nab\phi$, the potential $\phi$ is given by
\begin{equation}
\phi(\xx,t)={\cal N}\exp\{[\xx-\xx_{\rm f}(t)]^2/R^2\},
\end{equation}
where $R=2/\kf$ is the initial radius of the expansion waves and $\xx_{\rm f}(t)$
are random positions that change in forcing intervals $\delta t_{\rm f}$.
Here, the normalisation factor is ${\cal N}=\cs(\cs R/\delta t_{\rm f})^{1/2}$.

\section{Front speed versus gas speed}
\label{pposition}

\begin{figure}
\centering
\includegraphics[width=.6\textwidth]{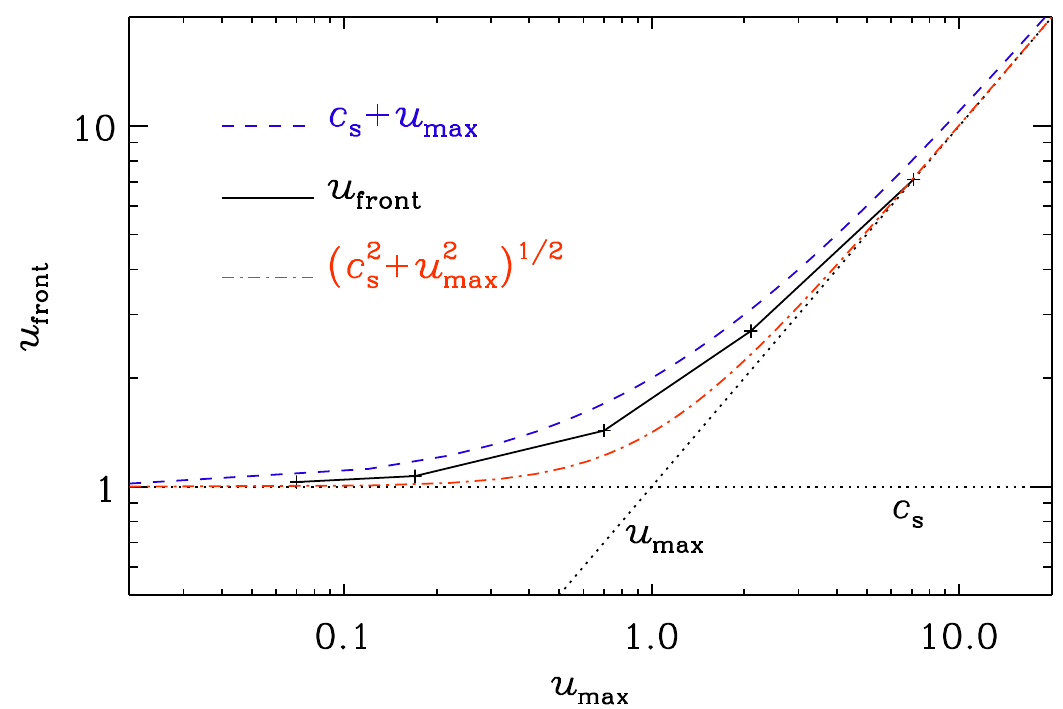}
\caption{
Front speed versus gas speed from the one-dimensional experiment described
in the text (plus signs and black line), compared with the associated
Doppler speed, $\cs+u_{\max}$ (blue) and $(\cs^2+u_{\max}^2)^{1/2}$ (red).
\label{fig:pposition}}
\end{figure}

In \Sec{Applicability}, we used counter-propagating acoustic waves to
drive inertial particles into each other.
The speed of these waves is equal to the sound speed when the speed is
small, but can become comparable to the gas speed for large velocities.
This is shown in \Fig{fig:pposition}.
Note that, even at subsonic gas speeds, the wave speed may exceed the
speed of sound.
Interestingly, the front speed is slightly faster than the associated
Doppler speed, $\cs+u_{\max}$, but slower than $(\cs^2+u_{\max}^2)^{1/2}$.

\bibliography{Refs.bib}
\end{document}